\documentclass[10 pt,final,journal,letterpaper,oneside,twocolumn]{IEEEtran}
\usepackage[utf8]{inputenc}
\usepackage[T1]{fontenc}
\usepackage{physics,amsmath}
\usepackage{amsfonts}
\usepackage{mathtools}
\usepackage{amsfonts}
\usepackage{amssymb}
\usepackage{makeidx}
\usepackage{csquotes}
\usepackage{docmute}
\usepackage{graphicx}
\usepackage{cite}
\usepackage{xcolor}
\usepackage{subcaption}
\usepackage{graphicx}
\usepackage{multicol}
\usepackage{multirow}
\usepackage{hyperref}
\usepackage[capitalise,nameinlink ,noabbrev]{cleveref}
\usepackage[algo2e,ruled, vlined, linesnumbered]{algorithm2e}
\begin{document}
\title{\huge{Joint Optimization of 3D Placement and Radio Resource Allocation for per-UAV Sum Rate Maximization}}
\author{\IEEEauthorblockN{Asad Mahmood, Thang X. Vu, \IEEEmembership{Senior Member,~IEEE}, Symeon Chatzinotas, \IEEEmembership{Fellow,~IEEE}, \\
Bj\"orn Ottersten, \IEEEmembership{Fellow,~IEEE}, }
\thanks{This work was supported by the Luxembourg National Research Fund via project 5G-Sky, ref. FNR/C19/IS/13713801/5G-Sky, and project RUTINE, ref. FNR/C22/IS/17220888/RUTINE. \newline
\indent {The part of this work has been accepted for publication at the IEEE VTC2023-Spring, Italy \cite{VTCFALL22}} \newline   
\indent The authors are with the Interdisciplinary Centre for Security,
Reliability and Trust (SnT), University of Luxembourg, 4365 Luxembourg
City, Luxembourg. Email: \{asad.mahmood, thang.vu, symeon.chatzinotas, bjorn.ottersten\}@uni.lu.}}
\maketitle

\begin{abstract}
Unmanned aerial vehicles (UAV) have emerged as a practical solution that provides on-demand services to users in areas where the terrestrial network is non-existent or temporarily unavailable, e.g., due to natural disasters or network congestion. In general, UAVs' user-serving capacity is typically constrained by their limited battery life and the finite communication resources that highly impact their performance. This work considers the orthogonal frequency division multiple access (OFDMA) enabled multiple unmanned aerial vehicles (multi-UAV) communication systems to provide on-demand services. The main aim of this work is to derive an efficient technique for the allocation of radio resources, $3$D placement of UAVs, and user association matrices. To achieve the desired objectives, we decoupled the original joint optimization problem into two sub-problems: (i) $3$D placement and user association and (ii) sum-rate maximization for optimal radio resource allocation, which are solved iteratively. The proposed iterative algorithm is shown via numerical results to achieve fast convergence speed after fewer than 10 iterations. The benefits of the proposed design are demonstrated via superior sum-rate performance compared to existing reference designs. Moreover, results showed that the optimal power and sub-carrier allocation help to mitigate the inter-cell interference that directly impacts the system's performance.
\end{abstract}
\begin{IEEEkeywords}
5G, multi UAV communication, convex optimization, internet of things (IoT).
\end{IEEEkeywords}
\section{Introduction}
\label{Intro}
\smallskip
\IEEEPARstart{T}he $5$G communication system aims to provide massive connectivity, ultra-reliable low-latency communication, and ensure the end-user quality of service. In contrast, according to  \cite{8744514,7510820}, terrestrial networks have not yet covered most remote areas due to the sparse human activities in these regions. Similarly, the goal of providing massive connectivity services is highly compromised in the case of an emergency where the existing communication system is disrupted due to natural disasters or faces congestion caused by the temporal increment in the number of users. Moreover, the excessive demand for the users' resources and connectivity represents a challenging task in meeting the objectives mentioned earlier \cite{8932190}. To overcome the above-mentioned challenges, different solutions have been proposed, e.g., device-to-device communication (D2D), to provide a wide range of communication services that allow devices in the vicinity of each other to communicate without relaying account the base station (BS). However, due to the adverse nature of wireless communication channels in emergency scenarios, it is challenging to optimize the route between the users \cite{8701700}. Second, satellite communication emerges as an effective solution for providing on-demand communication services to users in emergency scenarios, despite the low data rate and high latency that act as bottlenecks for widespread adoption, e.g., GEO satellites. \cite{9210567}. Furthermore, due to characteristics such as flexibility, adaptive altitude, and effortless deployment, unmanned aerial vehicles (UAVs) are emerging as an effective solution for providing reliable communication services to a user located in a particular location \cite{8932190}.
\par
UAVs, colloquially referred to as drones or flying base stations (BSs), have gained substantial attention in academia and industries, with applications ranging from photography to search and rescue operations, package delivery, agriculture, and other real-time applications \cite{7881122}. One of the promising but challenging tasks in UAV communication systems is determining how to deploy UAVs to provide massive coverage to user equipment (UEs) while meeting quality of service (QoS) requirements and improving network energy efficiency \cite{8918497}. Due to limited battery capacity, it should be carefully considered when designing the UAV deployment plan, together with other issues such as security, privacy, and flight regulations \cite{8660516}. {In multi-UAV systems, the placement problem is more challenging to guarantee a balanced trade-off between shadowing, path loss, and interference. Recent studies show that an increase in the altitude of UAVs leads to an increase in the probability of line-of-sight communication; in contrast, it also adds additional path loss \cite{7417609}. Hence, considering the joint UAV deployment in 3D cartesian coordinates and resource allocation is of great importance to enhance the UAV communication systems in emergency scenarios. Based on the above motivation, this work focuses on the joint 3D placement and resource allocation to improve the power efficiency of the multi-UAV network. The following subsection studies related works on UAV communication, followed by the main contributions of this paper.}

\subsection{Related Works}
The study on the UAV placement problem has recently been investigated in both $2$D \cite{9627548,8489918,8618602,9422153} and $3$D \cite{7918510,8698468} cartesian coordinates.
By employing a UAV as a relying node, the sum rate of the two-way relaying network can be improved via a joint optimization of the UAV trajectory and resource allocation while assuring the user QoS requirements \cite{9627548}. The authors of \cite{8489918} considered a UAV-enabled wireless power communication network to maximize common throughput by optimizing the trajectory and communication resources simultaneously while taking the UAV's maximum speed and users' energy neutrality into account. In \cite{8618602}, the authors maximize the UAV communication network's average secrecy rate by designing an optimal path and transmission power control.  
To improve QoS and provide communication services to edge users, the authors of \cite{9422153} considered an OFDMA-enabled UAV relay network. In \cite{7918510}, authors proposed a UAV placement algorithm to maximize served users using minimum transmission power. Furthermore, in \cite{8698468}, the authors considered a wireless sensor network scenario in which UAVs follow an optimal trajectory to maximize the data collection rate.
\par
The works above optimize the UAV deployment for a single-UAV network, which may not be applicable in high-demand scenarios. In such cases, employing multiple UAVs simultaneously could significantly improve the system performance if the UAV placement, user association, and transmission power control are properly designed \cite{8821282,8982086,8744514,9442809,8944019,7881122}.
In multi-UAV systems, 3D placement design has been proven to be more efficient in terms of the sum rate than 2D placement via optimizing the UAVs' altitude to maximize the UAV's coverage and minimize inter-UAV interference  \cite{8821282,8982086,8744514}.
Besides providing massive connectivity and ultra-reliable low-latency communication, energy efficiency and efficient utilization of resources are also essential performance metrics. 
To provide the same user demands, the smaller number of deployed UAVs is, the higher UAV utilization we have. 
The authors of \cite{9442809} proposed a bisection-based algorithm to determine the minimum number of UAVs among candidate locations which can provide service to all IoT users. It is noted that no bandwidth or power allocation was considered therein.
The authors of \cite{8944019} maximized the average capacity by deploying a minimum number of UAVs while ensuring that deployed UAVs can cover all massive or scattered users. A similar approach was considered in \cite{7881122}, which minimizes the number of UAVs in a 3D coordinate for different system scenarios.

{\subsection{Contributions}
\label{Contribution}
This work studies the performance of an OFDMA-enabled multi-UAV system deployed to provide on-demand communication services to ground users. OFDMA is regarded as a promising candidate for the beyond $5$G wireless networks due to its numerous benefits, such as flexible bandwidth and power allocation over users. \cite{6824752}. Therefore, it fits well into our UAV-enabled wireless communication network to maximize the system sum rate and minimize the number of deployed UAVs while meeting given QoS requirements through joint optimization of $3$D placement, sub-carrier allocation, and power control.
The most related works in the literature to ours are presented in \cite{9422153,8944019}. In \cite{9422153}, a single OFDMA-based UAV acts as the relay to provide services to cell-edge users of a BS. Whereas in \cite{8944019}, a 2D-based UAV placement algorithm was proposed to maximize the  average capacity of the UAV. It is noted that the solution in \cite{8944019} does not consider power and bandwidth allocation. Furthermore, \cite{8944019} considers only the free-space LoS channel model, while we consider the practically probabilistic channel model.
Our contributions are as follows:
\begin{enumerate}
    \item We consider the joint design of the 3D placement, sub-carrier allocation, and power control for the per-UAV sum-rate maximization in the downlink OFDMA-enabled multi-UAV communications to serve users in areas where the existing communication network is highly compromised by congestion or natural disaster. 
    \item We formulate a joint optimization of 3D UAV placement, user association, sub-carrier allocation, and transmit power to maximize the system sum rate while deploying a minimum number of active UAVs. Due to the inherent non-convexity, the original problem is decoupled into two sub-problems: i) path loss minimization and ii) sum rate maximization, which is then iteratively solved to provide the (local) optimal solution to the original joint problem.
    \item To solve the first sub-problem, we propose a heuristic algorithm to determine the minimum required UAVs based on the user demands and the ergodic capacity each UAV can provide, followed by an iteration algorithm to optimize the UAV 3D placement and user association. To tackle the non-convexity of the rate function in the second sub-problem, a successive convex approximation (SCA)-based iterative algorithm is proposed. In addition, a binary relaxation with a proper penalty function is implemented to accelerate the computation time. We show that the convergence to at least a local optimum of the proposed iterative algorithm is theoretically guaranteed. \item The effectiveness of the proposed approach is demonstrated by the numerical results, which show a significant improvement in the number of served users, power efficiency, and sum rate when deploying a minimum number of UAVs while adhering to the quality of service constraint of each user. 
\end{enumerate}}
\subsection{Organization}
The following sections summarize the rest of the paper: Section \ref{System Model and Problem Statement} summarizes the proposed system model and problem statement. Section \ref{Proposed Iterative Algorithm} presented an algorithm for efficient $3D$ UAV deployment and resource allocation. Likewise, Section \ref{Results and Discussion} contains the simulation results. The paper is finally concluded in Section \ref{Conclusion}

\section{System Model and Problem Formulation}
\label{System Model and Problem Statement}
\begin{figure}[htbp]
	\centering
	\includegraphics[width=0.8\linewidth]{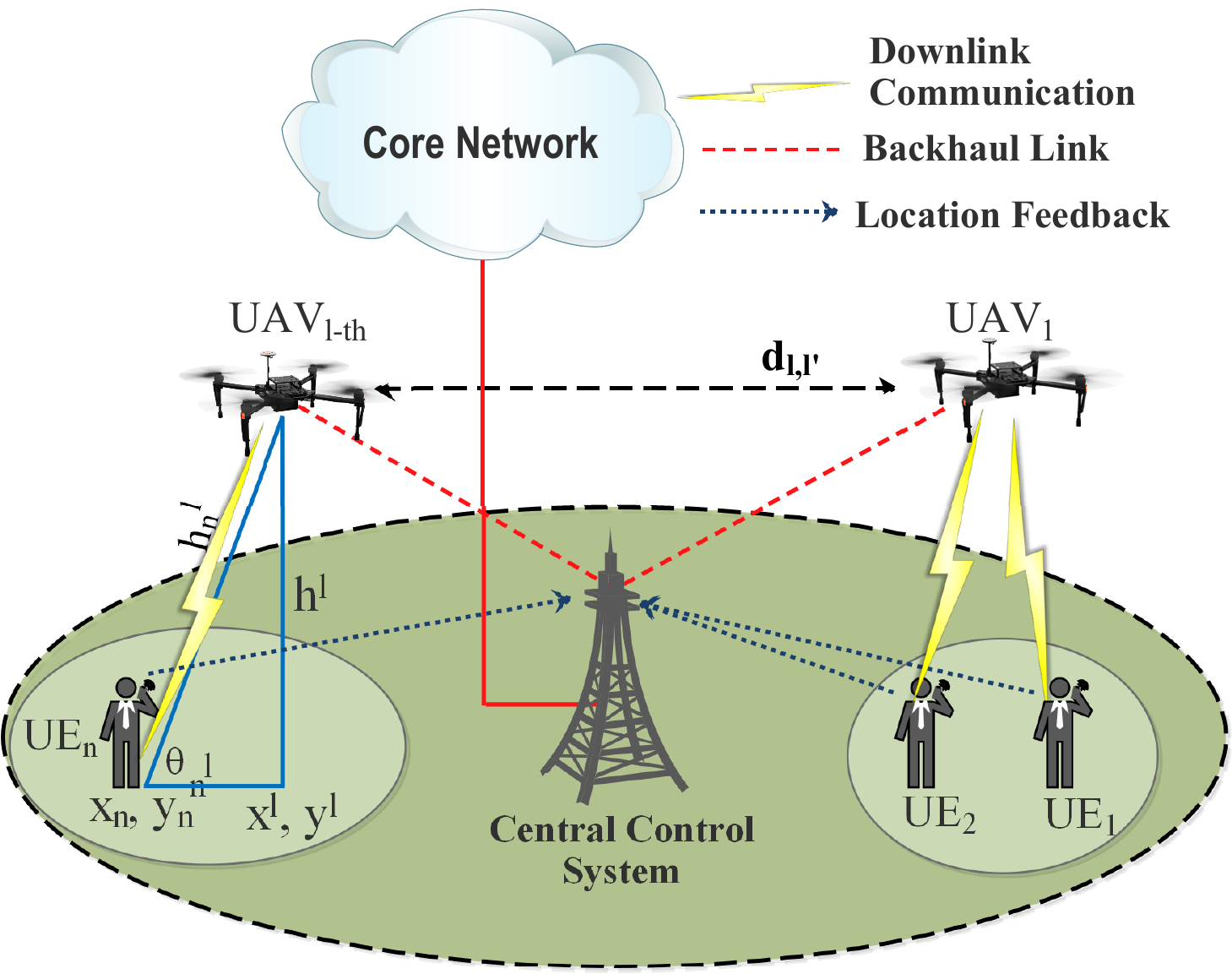}
	\caption{System model.}
	\label{fig:systemmodel1}
\end{figure}

In this work, we consider a down-link UAV-assisted wireless communication system that provides on-demand services to users in areas where the terrestrial network is non-existent or temporarily unavailable, e.g., due to natural disasters or network congestion, as shown in Fig.~\ref{fig:systemmodel1}. The considered system model consists of $M$ available UAVs equipped with $U$ antennas and $N$ users, which are represented by the sets $\mathcal{M} = \{ 1,2, \cdots {\rm{M\} }}$ and $\mathcal{N} = \{ 1,2, \cdots {\rm{N\} }}$, respectively. In order to improve UAVs' deployment usage, only $L\leq M$ UAVs will be activated to serve the users. The set of active UAVs is denoted by $\mathcal{L} \subset \mathcal{M}$. We consider $3D$ coordinates, in which the location of the $l$-th UAV is represented by $(x^l,y^l,h^l)$, which satisfy:
\vspace{-1mm}
\begin{equation}
\small
\label{Bounds}
    \left\{ {\begin{array}{*{20}{c}}
x_{min} \le {x^l} \le {x_{\max }}\},{\forall l} \\
y_{\min} \le {y^l} \le {y_{\max }}\},{\forall l}\\
h_{\min} \le {h^l} \le {h_{\max }}\},{\forall l}
\end{array}} \right..
\end{equation}
where $(.)_{min}$ and $(.)_{max}$ represent the lower and upper bound of $l$-th UAV coordinates, respectively. The coordinates of the users are denoted by $\{x_n, y_n, 0\}$. For notation convenience, we denote ${z^l} \triangleq \{{x^l},{y^l}\}$ and ${v_n} \triangleq  \{{x_n},{y_n}\}$ as the horizontal coordinates of $l$-th UAV and $n$-th UE. The UAVs are assisted by a central controller that provides information about UE's location\cite{8918497}. The total users are divided into clusters, each of which will be served by one active UAV. In order to minimize interference, each UAV serves its connected users via  OFDMA \cite{9422153}, which consists of $K$ sub-channels denoted by $\mathcal{K}=\left\{ {1, \cdots ,K} \right\}$. We note that although there is no intra-cluster interference, there exists inter-cluster interference caused by the transmission of neighbouring UAVs.

\par
Let us denote the sub-channel allocation matrix by $\textbf{A}\!\! =\!\! [a_n^{k,l}{{\rm{]}}_{N\! \times\! K\! \times\! L}}$, where $a_n^{k,l}=1$ indicates that sub-channel $k$ served by UAV $l$ is allocated to user $n$ and  $a_n^{k,l}=0$ otherwise.
In addition, we denote $p_n^{k,l}$ as the transmit power from UAV $l$ to user $n$ over sub-channel $k$. Since UAV $l$ only consumes power on sub-channel $k$ for user $n$ if that sub-channel is assigned to user $n$, we introduce the following constraint:

\begin{equation}
\small
 \label{Power allocation}
    {p_n^{k,l}\le a^{k,l}_n P_t^{Max}}, p_n^{k,l}\ge 0,a_n^{k,l} \in \{0,1\} ,\forall n,k,l.
\end{equation}
where $P_t^{Max}$ is the maximum transmit power of the UAV. 
The transmit power constraint \eqref{Power allocation} not only facilitates the rate function, as will be shown later in this section but also guarantees no transmit power on sub-channel $k$ if it is not assigned to user $n$. To avoid mutual interference among the users, each sub-channel in one cluster (served by one UAV) is allocated to only one user, which imposes the following constraint:
\begin{equation}
\small
\label{Subcarrier 1}
    {\sum}_{n \in \mathcal{N}} {{\rm{a}}_n^{k,l}}  \le 1,\forall k,l.
\end{equation}
\par
Similarly, the user association $\textbf{J}\!\!=\!\![J_n^l]_{N\!\times L}$ is a binary matrix, in which each entity $J_n^l$ represents the connection between $l$-th UAV and $n$-th UE, i.e., $J_n^l=1$ if user $n$ is associated with UAV $l$, and $J_n^l=0$ otherwise. Consequently, the following constraint must be met as one subchannel is only allocated to a user if it is connected to the corresponding UAV:
\vspace{-1mm}
\begin{equation}
\small
\label{Subcarrier 2}
    {\sum}_{k \in {\cal K}} {{\rm{a}}_n^{k,l}}  \le J_n^l,\;\; J_{n}^l \in \{0,1\},\forall n,l.
\end{equation}
\vspace{-1mm}
\subsection{Path Loss Model}
We consider the general probabilistic channel modelling in which the channel between UAVs and their associated users includes both LOS and non-LOS (NLOS) parts. The probability of LoS is primarily determined by the 3D geographical location of UAVs and users on the ground, as well as the surrounding environment, as given below \cite{7918510}:
\vspace{-2mm}
\begin{equation}
\small
\label{Prob_LoS}
   PLos_n^l = \frac{1}{1+{b_1}\exp \left( { - {b_2}(\theta _n^l - {b_1})}\right)}. 
\end{equation}
In \eqref{Prob_LoS}, $b_1$ and $b_2$  are the constants representing the environment condition, and $\theta _{n}^l={180}/\left(\pi \sin\left(h_l/d_n^l\right)\right)$ is the angle of elevation between the $l$-th UAV and the $n$-th UE,

where $d_n^l=\sqrt{||z^l-v_n||^2+{h^l}^2}$ is the distance between the $l$-th UAV and $n$-th user. The probability of non-line of sight (PNLos) communication is represented by $PNLos_n^l = 1 - PLos_n^l$. 

The air-to-ground (A2G) path loss model of both LoS and NLoS  between $l$-th UAV and $n$-th user is given as
\begin{equation}
\small
\label{Loss Model}
\hat{PL_n^l}=\left\{ {\begin{array}{*{20}{c}}
\xi_{LoS}K_o,\;\;for\;\;LoS \\
\xi_{NLoS}K_o,\;for\;\;NLoS
\end{array}} \right.,
\end{equation}
where $\xi_{NLoS}$ and $\xi_{LoS}$ represent the attenuation factors for both NLoS and LoS links, respectively; $K_o = {\left( {\frac{{4\pi {f_cd_n^l}}}{c}} \right)^\alpha}$; $f_c$, $\alpha$ and $c$  respectively represent the carrier frequency, path loss exponent and speed of light. Therefore, the UAV-User path loss is given as follows \cite{8727504}:
\begin{equation}
\small
\label{channel}
{PL}_n^{l}=K_o\left[PLos_n^l\xi_{LoS}+{\xi_{NLoS}}PNLos_{n}^l\right].
\end{equation}
{\subsection{Transmission Model}
\label{Transmission_Model}
Let $\boldsymbol{h}^{k,l}_n \in \mathbb{C}^{U\times 1}$ be the small-scale fading from UAV $l$ to user $n$ on sub-channel $k$. To improve the receive SNR, we employ the maximum ratio transmission preceding, which results in an effective channel gain $g _n^{k,l}=|( \boldsymbol{h}_n^{k,l})^H \boldsymbol{h}_n^{k,l} |^2/{PL}_n^{l}$, where ${PL}_n^{l}$ is given in \eqref{channel}. transmit ratio The downlink signal received on the $k$-th subchannel at user $n$ from UAV $l$ is given as follows:
\vspace{-1mm}
\begin{equation}
\small
\label{trasmission signal}
    y_n^{k,l}= \underbrace {\sqrt{p_n^{k,l}}g_n^{k,l}s_n^{k,l}}_{\text{Information signal}} + \underbrace {\sum_{l' \ne l} {\Big( {\sum_{n' \ne n}\sqrt{p_{n'}^{k,l'}}} \Big)} g_n^{k,l'}s_{n'}^{k,l'}}_{\text{Inter-cell interference}}+ n_n^k.
\end{equation}
where $p_n^{k,l}$ and $s_n^{k,l}$ represent the transmission power of $l$-th UAV and information transmitted signal with $\mathbb{E}|s_n^{k,l}|^2= 1$, respectively, and $n_n^k$ is the Gaussian noise with zero mean and variance $\sigma^2$.
 The first term on the right-hand side of \eqref{trasmission signal} represents the information signal transmitted over the sub-carrier $k$ by the $l$-th UAV to the $n$-th user. The second term represents the interference caused by other UAVs' transmission on the same sub-carrier $k$. Therefore, the signal-to-interference plus noise ratio (SINR) of sub-channel $k$ of at user $n$ from UAV $l$ is given as
\begin{equation}
 \label{SINR}
\gamma{_n^{k,l}} = {p_n^{k,l}g_n^{k,l}}/(\Phi _n^{k,l}  + {\sigma ^2}),
 \end{equation}
where  $\Phi _n^{k,l} =\sum_{l \ne l'} {\left( {\sum_{n \ne n'}p_{n'}^{k,l'}} \right)} g_n^{k,l'}$ is the aggregated interference imposed by other UAVs. Thus, the achievable rate of $n$-th UE connected to $l$-th UAV is calculated as
\begin{equation}
    \label{Rate carrier}
{R_n^{l}} = {B}{\sum}_{k \in \mathcal{K}}{\log _2}\left( {1 + \gamma{_n^{k,l}} } \right).
\end{equation}}
\subsection{Problem Formulation}
Unlike previous works, which target only the sum-rate maximization, we also optimize the system-wise energy usage. Therefore,  our objective is to maximize the ratio between the system sum rate and the number of active UAVs, which is given as $\Upsilon  = \frac{{{R_{sum}}}}{L} = \frac{\sum_{n,l}R_n^l}{L}$, where $R_n^l$ is calculated in \eqref{Rate carrier}, and $L$ is the number of active UAVs. We aim to jointly optimize the UAVs' 3D placement, user association, power and subcarrier allocation to maximize the objective function $\Upsilon$. 
For ease of presentation, we introduce short-hand notations $\textbf{W} = \left[ {W^l\triangleq {x^{l}},{y^{l}},{h^{l}}} \right]$, $\textbf{P}= [{p_{n}^{k,l}}]{_{N \times K \times L}}$, $\textbf{J}= [{J_{n}^l}]{_{N \times L}}$, and ${{\textbf{A} = [a}}_n^{k,l}{{\rm{]}}_{N \times K \times L}}$. Furthermore, let us denote $r_{0n}$ as the QoS requirement of user $n$, and $C^l_{Min}, C^l_{Max}$ as the minimum and maximum numbers of users the UAV can serve, respectively. Then the joint optimization is formulated as follows:
\begin{subequations}
	\label{Main Equation1}
	\begin{align}
		&\mathbb{P}_0:\mathop {\max }_{L,\textbf{W},\textbf{P},\textbf{J},\textbf{A}} \frac{1}{L}{\sum}_{l\in \mathcal{L}} {\sum}_{n\in \mathcal{N}}R_n^l \\
		&\text{s.t.}\;\text{\cref{Bounds,Power allocation,Subcarrier 1,Subcarrier 2}}.\\
	\label{C1}	&\;\;\;\;R_n^l \ge J_n^l r_{0n}, \forall n,l ,\\ 
	\label{C2}
	&\;\;\;\;{\sum}_{l\in \mathcal{L}} {J_{n}^l} \le 1 ,\forall n, \\
	\label{C3_4}
	&\;\;\;\;{\sum}_{{n} \in \mathcal{N}} {{J_{n}^l} \in [C^l_{Min}, C^l_{Max}]} ,\forall l,\\
	\label{C5}
	&\;\;\;\;{\sum}_{n \in \cal{N}} {{\sum}_{k \in \mathcal{K}} {p_n^{k,l} \le P_t^{Max}} },\forall l ,  \\
	\label{C6}
	&\;\;\;\;{\sum}_{n \in \cal{N}} {{\sum}_{l \in \cal{L}}}J_n^l  \ge \lambda N ,\\
		\label{C7} 
	&\;\;\;\;\mathop {\| {{W^l} - {W^{l'}}} \|} \ge {d_{_o}},{\forall l}.
    \end{align}
\end{subequations}
Our objective in this work is to deploy a minimum number of UAVs to maximize the average serving rate, subject to QoS and system resource requirements. {In problem $\mathbb{P}_0$, constraint \eqref{C1}  ensures the QoS requirement of corresponding UE-UAV link}; 
\eqref{C2} guarantees that no UE is connected to more than one UAV; \eqref{C3_4} represents the bounds of the serving capacity of each UAV;  
\eqref{C5} restricts the total transmit power at each UAV should not exceed the maximum power; \eqref{C6} guarantees the minimum $\lambda$ per cent of the UEs should be served; finally, \eqref{C7} constraints the distance between the two neighbouring UAVs not exceeding the minimum allowable distance to avoid a collision. {Note that by setting $\lambda$ equals to 1, constraints (11g) together with (11c) assure the QoS requirements for all users.}
\par
\section{Framework for  Optimal 3D Placement, User Association, and Radio Resource Allocation}
\label{Proposed Iterative Algorithm}
\begin{figure}[ht]
	\centering
	\includegraphics[width=0.95\linewidth]{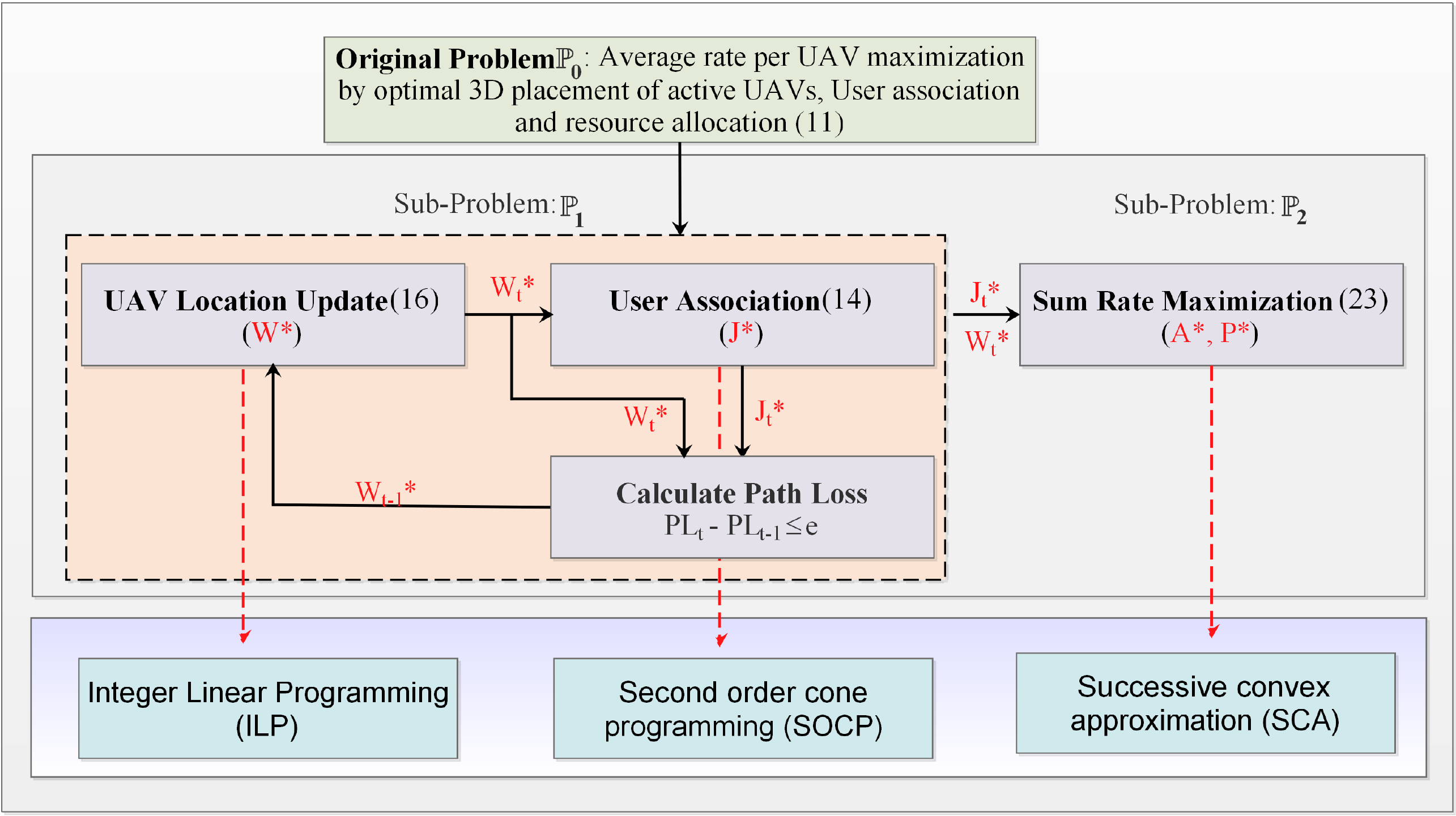}
	\caption{Road-map of iterative algorithms.}
	\label{fig:RoadMap}
	\vspace{-0.3cm}
\end{figure}

\vspace{-1mm}
{
The optimization problem in \eqref{Main Equation1} is mixed-integer non-linear programming (MINLP) and NP-hard due to the integer nature of the user association matrix $\textbf{J}$, and sub-carrier allocation matrix $\textbf{A}$. Furthermore, it is challenging to get optimal results as it involves non-linear, non-convex objective function and constraints. Towards an efficient solution to problem $\mathbb{P}_0$ in \eqref{Main Equation1}, we decouple the original optimization problem into two sub-problems: i) the first sub-problem, denoted by $\mathbb{P}_1$, optimizes the number of active UAVs, their optimal 3D placement, and the user association to minimize the average path loss;
ii) and the second sub-problem, denoted by $\mathbb{P}_2$, optimizes the radio resources (power and subcarrier allocation) 
to minimize inter-cell interference and maximize the sum rate of users.}
In the following, we will detail the sub-problems, whose overview diagram of the problem decomposition is shown in Fig.~\ref{fig:RoadMap}.

{\subsection{Solving $\mathbb{P}_1$: Optimal 3D Placement and User Association}
\label{Proposed Iterative Algorithm PL}
The first sub-problem is directly obtained from the original problem by treating the radio resources ($\boldsymbol{A}$ and $\boldsymbol{P}$) as constants, which is formulated as follows:
	\begin{align}
		&\mathbb{P}_1:\mathop {\max }_{L,\textbf{W},\textbf{J}} ~~\frac{1}{L}{\sum}_{l\in \mathcal{N}} {\sum}_{n\in \mathcal{N}}R_n^l \\
		&\text{s.t.}\; \text{\cref{Bounds,Subcarrier 2,C1,C2,C3_4,C6,C7}}. \notag
    \end{align}
Jointly optimizing the number of active UAVs, their 3D locations, and user association is challenging. To address this issue, we initially present a practical and effective heuristic approach to determine the optimal number of required UAVs denoted by $L$. Intuitively, $L$ should depend on factors such as the number of ground users, the QoS requirements, and the average capacity of the networks \cite{8833519}. By employing Shannon's capacity bound, the average spectral efficiency of any arbitrary user can be calculated as follows:
\begin{equation}
\label{Shannon Bound}    \Lambda=\mathbb{E}\Big[\log_2\Big(1+\frac{P_t^{max}}{f_c\sigma^2}\Big(\frac{|\textbf{h}^H\textbf{h}|^2}{{PL}}\Big)\Big)\Big].
\end{equation}
Based on equation \eqref{channel}, the path loss is primarily determined by the LoS channel, which is dependent on the 3D geographical locations of the UAVs. We assume that the UAVs are positioned in a manner that ensures LoS dominance, i.e., $PLos \simeq 1$. Hence, the maximum data rate that a user can achieve over the bandwidth $B$ is represented by $R^{max}=B \times \Lambda$. To satisfy the QoS requirement $r_{0n}$, the maximum number of user equipment (UEs) that each UAV can support is given by ${C^{\max}} = \left\lfloor \frac{R^{max}} {r_{0n}} \right\rfloor, \forall n$. Consequently, we employ a heuristic approach to determine the number of active UAVs denoted by $L = \left\lceil {\frac{ \lambda N}{{{C^{\max}}}}} \right\rceil$, where $\lambda$ denotes the percentage of users that need to be served. Let $\mathcal{L}$ represent the set of active UAVs.
\\
 Once the number of active UAVs $L$ is determined, the sub-problem $\mathbb{P}_1$ focuses on optimizing the placement of active UAV's in 3D space and their user associations. The altitude of each UAV is a critical factor in determining the probability of the line-of-sight (LoS) channel $PLos_n^l$. As the altitude of the UAV increases, so does the angle of elevation $\left(\theta_n^l\right)$, resulting in an increase in the value of $PLos_n^l$. However, higher altitude also leads to greater path losses between the UAV and its associated users, resulting in reduced sum rate \cite{7037248}. Thus, the primary objective of the first sub-problem is to minimize path losses. To achieve this goal, the UAVs should be placed in locations where the probability of LoS communication is greater than or equal to a threshold value $\phi$, i.e., $PLos_n^l \left(\theta_n^l\right) \ge \phi$. Additionally, each user should only be associated with the UAV located at a distance less than $h^l/\sin(PLos_n^l{^{-1}}(\phi))$.}
\subsubsection{User Association}
\label{User_Association}
Based on the above analysis, we propose to solve the first sub-problem via an alternative optimization algorithm, which treats each of the UAVs' locations and the user association matrix independently, assuming the other is fixed. Therefore, we model an iterative mathematical framework for optimal $3$D placement and user association to minimize the path loss between the UAVs and their associated users.
Following that, the UEs association problem, given the location of UAVs and UEs, can be formulated as follows:\vspace{-0.3mm}
\begin{subequations}
	\label{Main Equation2a}
	\begin{align}
		\mathbb{P}_1^a:&\; \mathop {\min }_{\bf{J}} {\sum}_{n \in \cal{N}}{{\sum}_{l \in \cal{L}}{{J_n^l}{{\left\| {{{\bf{V}}_n} - {{\bf{W}}_l}} \right\|}^2}} } \\
~&~\text{\cref{C2,C3_4,C6}},\\
	 &{J_n^l}\left\| {{{\bf{V}}_n} - {{\bf{W}}_l}} \right\| \le \frac{{{h_l}}}{\sin(PLos_n^l{^{-1}}(\phi))},\forall n,l.
	\end{align}
\end{subequations}
where ${{\bf{V}}_n} \triangleq\left( {{x_n},{y_n},0} \right)$ represents the location of user $n$. The optimization problem mentioned in \eqref{Main Equation2a} is integer linear programming and convex in nature and can be solved using the standard methods, e.g., integer linear programming (ILP).
\subsubsection{UAV's Location}
\label{UAV_Location}
Given the user association matrix $\textbf{J}$ calculated in problem \eqref{Main Equation2a}, the location of the UAV is updated such that the distance between the UAV and associated UEs is minimized. Denote $Q^l$ $\in$ $\mathcal{Q}$ as the number of UEs associated with the $l$-th UAV, the UAVs'$3$D placement problem can be formulated as follows:
\begin{subequations}
	\label{Main Equation2b}
	\begin{align}
	\label{P2b Objective}
		\mathbb{P}_1^b:&\mathop {\min }_{\mathbf{W}} {\sum}_{n \in {\cal{Q}}^l}{\left[ {{{\left( {{x_n} - {x_l}} \right)}^2} + {{\left( {{y_n} - {y_l}} \right)}^2} + h_l^2} \right],\forall l},  \\
		\label{P2b Constraint}
   &{\left( {{x_n} - {x_l}} \right)^2} + {\left( {{y_n} - {y_l}} \right)^2} + h_l^2\xi  \le {\rm{0, }}\forall n \in {\cal{Q}}^l,\\
&\text{\cref{Bounds,C7}}.
		\end{align}
\end{subequations}

The parameter $\xi=1-1/\sin(PLos_n^l{^{-1}}(\phi))$ is introduced in constraint \eqref{P2b Constraint} to ensure that the UAV is placed at the point where the angle of elevation calculated based on its geographical location should satisfy $PLos_n^l \left(\theta_n^l\right) \ge \phi$. Because the objective function and the first constraint are both quadratic, hence \eqref{Main Equation2b} is a quadratic constraint quadratic programming problem (QCQP).
\textit{Lemma 1}: Problem \eqref{Main Equation2b} can be equivalently formulated as a following QCQP problem:
\begin{subequations}
	\label{Main EquationP2b3}
	\begin{align}
	\label{EquationP2b2 Objective}
	\mathbb{P}_1^{b,1}&:\mathop {\min }_{\textbf{W}} \frac{1}{2}{{\textbf{{W}}}^T}{\textbf{H} _o}\textbf{W} + \mathbf{F}_o^T{\textbf{W} }+{\kappa _{\rm{o}}},\forall l  \\
		\label{EquationP2b2 Constraint}
	&\frac{1}{2}{{\bf{W}}^T}{\textbf{H}_n} + \mathbf{F}_n^T{{\textbf{W} + }}{\kappa _n} \le 0,\forall n \in {\cal{Q}}^l.
		\end{align}
\end{subequations}
Where
\begin{equation}
\small
\label{convexity}
 {\textbf{H} _o}= \begin{bmatrix}
2Q^l&0&0\\
0&2Q^l&0\\
0&0&2Q^l
\end{bmatrix},  {\textbf{H} _n}= \begin{bmatrix}
2&0&0\\
0&2&0\\
0&0&\xi
\end{bmatrix}.
\end{equation}

\begin{equation}
 {\textbf{F} _o}\!\!=\!\! \begin{bmatrix}
-2\!\!\!\sum\limits_{n\in Q^l}\!\!\!{x_n}&\!\!\!\!\!-2\!\!\!\sum\limits_{n\in Q^l}\!\!\!{y_n}&0
\end{bmatrix}^T\!\!,  {\textbf{F} _n}\!\!=\!\! \begin{bmatrix}
-2x_n&0y_n&0
\end{bmatrix}^T.
\end{equation}
\begin{equation}
    \kappa_o={\sum}_{n\in Q^l}{x_n^2}+{\sum}_{n\in Q^l}{y_n^2}, \;\;\; \kappa_n=x_n^2+y_n^2.
\end{equation}
\begin{IEEEproof}
See Appendix~\ref{Lemma1}.
\end{IEEEproof} 
Solving a QCQP optimization, in general, is challenging because of its NP-hard nature \cite{lu2011kkt}. Fortunately, there is a subclass of QCQP problems in which both the objective function and constraint are convex in nature, as demonstrated in \eqref{convexity}. This subclass of problems can be efficiently solved using the second-order cone programming method \cite{lobo1998applications}.
\par
By iteratively solving problems \eqref{Main EquationP2b3}  and \eqref{Main Equation2a}, the average path loss and the user association can be optimized, whose steps are listed in Algorithm~\ref{algo:Algorithem 1}.
\begin{algorithm2e}[t]
\small
\SetAlgoLined
\textbf{Initialization: } $PL_{old}$, error\;
\textbf{Execution: }\;
\While{error $>$ $\epsilon$}{
        Given the random location of UAV's Solve \eqref{Main Equation2a} Using Branch and Bound (BnB) algorithm  to get the User association matrix $\textbf{J}$.\\
        Based upon the User Association matrix, update the location of UAV by solving \eqref{Main Equation2b}\\
        Compute Path Loss using using \eqref{channel} to obtain $PL_{*}$ \\
        Compute \textbf{error} = $|PL_{*}$-$PL_{old}|$\\
        Update $PL_{old}$=$PL_{old}-PL_{*}$
        
}
\caption{\textbf{\small{Iterative Algorithm to solve} \eqref{Main Equation2a}  and \eqref{Main EquationP2b3}}}
\label{algo:Algorithem 1}
\end{algorithm2e}

\subsubsection{Convergence and Complexity Analysis of Algorithm \ref{algo:Algorithem 1}}
\paragraph{Convergence Analysis}
\begin{equation}
\small
\label{Convergence}
    \mathbb{F}(\textbf{W}^{t+1}) \le \mathbb{F}(\textbf{W}^t), \;\;     \mathbb{F}(\textbf{J}^{t+1}) \le \mathbb{F}(\textbf{J}^t).
\end{equation}
Equation \eqref{Convergence} perceives that the objective function values of \eqref{Main Equation2a} and \eqref{Main EquationP2b3} decrease as with the iterations; as a result path loss $PL(\mathbf{W, J})$ starts decreasing and converges to a finite value when the difference between the current and previous iteration is less than epsilon $\epsilon$, i.e., $PL^{t}-PL^{t-1}\le \epsilon$. 
\paragraph{Complexity Analysis}
This section provides the per iteration worst-case complexity analysis of Algorithm \ref{algo:Algorithem 1} that solves \eqref{Main Equation2a} and \eqref{Main EquationP2b3} iteratively. Since the \eqref{Main Equation2a} is the least square minimization problem and convex in nature, it can be solved efficiently using integer programming. More specifically, a convex problem \eqref{Main Equation2a} involves ${NL}$ decision variables and $2{NL}+2{L}+2{L}+1$ constraints. Because of this, we can express $\left(NL\right)^3\left(2{NL}+2{L}+2{L}+1\right)$ as the complexity required to solve \eqref{Main Equation2a} in each iteration. On the other hand, in optimization problem \eqref{Main EquationP2b3}, compromise of $3Q^l$ decision variables and $Q^l+7L$ constraints. Likewise,
to solve \eqref{Main EquationP2b3}, the required per-iteration complexity is $\left(3Q^l\right)^{0.5}\left(Q^l+7L\right)$. Furthermore, assuming that $t_{m}$ are a total number of iterations, the total complexity of Algorithm \ref{algo:Algorithem 1} is $\mathcal{O}\left(t_{m}\!\left(\!\left(NL\right)^3\!\left(2{NL}\!\!+2{L}+\!2{L}\!+1\right)\!+\!\left(3Q^l\right)^{0.5}\!\!\left(Q^l+7L\right)\right)\!\!\right)$.

\subsection{Solving $\mathbb{P}_2$: Joint Optimization of Transmit Power and Subcarrier Allocation}
\label{JTPSCA}
Once the UAVs' location and the user association matrix are defined, the second subproblem will jointly optimize the transmit power at the UAVs and the subcarrier allocation to maximize the sum rate while guaranteeing users' QoS. Mathematically, the second subproblem is formulated as follows:
\begin{subequations}
	\label{Main Equation3}
	\begin{align}
	\mathbb{P}_2^{a}:\;\;& \mathop {\max }_\textbf{P, A} {\sum}_{n \in \mathcal{N}} {{\sum}_{l \in \mathcal{L}} R_n^l}\\
	\label{Main Equation3C1}
	&{R_n^l \ge J_{n}^lr_{0n}},\forall n\in \mathcal{N}, l\in \mathcal{L}, \\
	&\text{\cref{C5,Power allocation,Subcarrier 1,Subcarrier 2}}.
		\end{align}
\end{subequations}
where $R^l_n$ is given in \eqref{Rate carrier}. The problem \eqref{Main Equation3} is non-convex in nature due to $R^l_n$ in objective function and constraint \eqref{Main Equation3C1}. To tackle this difficulty, we introduce auxiliary variables $\eta_n^l, s_n^{k,l}, \forall n,l$ and reformulate \eqref{Main Equation3} as 
\begin{subequations}
	\label{Transformed2}
	\begin{align}
	\label{Convex Equation Objective 2}
	\mathbb{P}_2^{a,1}:&\;\mathop {\max }_{{\bf{A}},{\bf{P}},{\bf{\eta }},{\mathfrak{s}}} {\sum}_{n \in N} {{\sum}_{l \in L} {\eta _n^l} } \\
	\begin{split}
	\label{T2C1}
	&\;{\sum}_{k \in \mathcal{K}} \log \left(\Phi_n^{k,l}+\delta^2+p_n^{k,l}g_n^{k,l}\right) \ge \frac{{\log (2)\eta _n^l}}{B}\\
	&\qquad+ {\sum}_{k \in \mathcal{K}}s_n^{k,l}, \forall n \in \mathcal{N}, l \in \mathcal{L} ,
	\end{split}\\
	\label{C3T3}
	&\;\Phi_n^{k,l}+\delta^2 \le e^{s_n^{k,l}}, \forall n,l,k,\\
	&\;\eta^l_n \ge J^l_n r_{0n}, \forall n \in \mathcal{N}, l \in \mathcal{L} \label{C3T2}, \\
	&\text{\cref{C5,Power allocation,Subcarrier 1,Subcarrier 2}}.
	\end{align}
\end{subequations}
where \eqref{C3T2} guarantee the QoS requirement, and \eqref{T2C1} and \eqref{C3T3} equivalently represent $R_n^l \geq \eta_n^l$.  
Solving \eqref{Transformed2} is still challenging due to the non-convex constraint \eqref{C3T3}. Fortunately, it has a form of different-of-convex (DC) constraint; hence we can employ the first-order approximation to convexify the right-hand-side of \eqref{C3T3}, which can be reformulated as follows:\vspace{-2mm}
\begin{subequations}
	\label{Transformed3}
	\begin{align}
	\label{Convex Equation Objective 3}
	\mathbb{P}_2^{a,2}:&\;\mathop {\max }_{{\bf{A}},{\bf{P}},{\bf{\eta }},{\mathfrak{s}}} {\sum}_{n \in N} {{\sum}_{l \in L} {\eta _n^l} } \\
	\label{C3T4}
	&\;{\Phi_n^{k,l}+\delta^2\le e^{{s_o}_n^{k,l}}\left(s_n^{k,l}-{s_o}_n^{k,l}+1\right)} \notag \\
	&\qquad\qquad \qquad\qquad\forall n \in \mathcal{N},k\in\mathcal{K},l\in\mathcal{L}, \\
	&\text{\cref{C5,T2C1,C3T2,Power allocation,Subcarrier 1,Subcarrier 2}}.
	\end{align}
\end{subequations}
where ${s_o}_n^{k,l}$ is a feasible point of constraint \eqref{C3T4}.
Since the objective is linear and all constraints are either convex or linear with binary variables $a^{k,l}_n$, problem \eqref{Transformed3} can be solved by the standard branch-and-bound method via, e.g., Mosek solver. This method, however, incurs exponential complexity with the problem size. Therefore, we relax the binary variables into continuous ones. In addition, a penalty function $\Psi_n^l={\sum}_{k \in \mathcal{K}}((a_n^{k,l})^2-a_n^{k,l})$ is added to the original objective function to promote the convergence. 
However, since $\Psi_n^l$ is convex, it cannot be added to the objective function. In order to accomplish this, we employ its first-order approximation, computed as
\begin{equation}
    \label{Penalty 1}
    {\Psi}_n^l={\sum}_{k \in \mathcal{K}} \left(2a_{o,n}^{k,l}a_n^{k,l} -(a_{o,n}^{k,l})^2  - a_{o,n}^{k,l}\right).
\end{equation}
By using \eqref{Penalty 1} as the approximated penalty function, a relaxed resource allocation optimization problem with penalty parameter $\mu$ can be formulated as follows:
\begin{subequations}
	\label{Transformed3_relaxed}
	\begin{align}
	\label{Convex Equation Objective 3_R}
	\mathbb{P}_2^{b}:&\;\mathop {\max }_{{\bf{A}},{\bf{P}},{\bf{\eta }},{\mathfrak{s}}} {\sum}_{n \in N}{\sum}_{l \in L}\left(\eta_n^l+\mu\Psi_n^l\right) \\
	&\;0\le a_n^{k,l} \le 1,\\
	&\text{\cref{C5,Power allocation,Subcarrier 1,Subcarrier 2}},\\
	&\text{\cref{T2C1,C3T2,C3T4}}.\nonumber
	\end{align}
\end{subequations}
It is observed that problem \eqref{Transformed3_relaxed}, which is the relaxed version of \eqref{Transformed3}, is convex in nature because of the linear objective function and convex constraints. Thus, problem \eqref{Transformed3_relaxed} can be solved efficiently using standard methods, e.g., the interior point. We note that problem \eqref{Transformed3_relaxed} provides a (sub) optimal solution to problem \eqref{Transformed3} since it satisfies all constraints of this problem. In order to narrow the gap to the optimal solution of problem \eqref{Transformed3}, an iterative Algorithm \ref{algo:Algorithem 2}, constitute a sequence of convex optimization problems, is proposed. The key idea behind Algorithm \ref{algo:Algorithem 2} is to have better values of feasible variables in the approximated optimization problem. 
\begin{algorithm2e}[t]
\small
\SetAlgoLined
\textbf{Initialization: } $s_o$, $\eta_{old}$, error\;
\textbf{Execution: }\;
\While{error $>$ $\epsilon$}{
        Given the location of UAVs and user association matrix form Algorithm \ref{algo:Algorithem 1}, Solve \eqref{Transformed3} to get the optimal value of $\eta _*$,$a _*$, $p_*$,$s_*$ \\
        Compute \textbf{error} = $|\eta_{*}$-$\eta_{old}|$\\
        Update $\eta_{old}$=$\eta_{*}$,$s_{o}$=$s_{*}$
        
}
\caption{\textbf{{Iterative \!Algorithm to\! solve\!} \eqref{Main Equation3}}}\label{algo:Algorithem 2}
\end{algorithm2e}

\begin{table}[t]
    \centering
    {\color{black}
    \caption{Environment values. H-RB:\! High-Rise Buildings \cite{7510820}}
    \label{tab:modeling parameters}
    \begin{tabular}{c c c c c}
    \hline \hline
    Environment&Sub-Urban & Urban &Dense-Urban & H-RB \\ \hline \hline
$\xi_{LoS}$&1&1&1.6& 2.3\\
$\xi_{NLoS}$& 21& 20&23&34\\
$b_1$&4.88&9.61&12.08&27.23\\
$b_2$&0.43& 0.16&0.11&0.08\\
\hline
    \end{tabular}}
\end{table}

\begin{table}[t]
    \centering
    {\color{black}
    \caption{Simulations Parameters}
    \label{tab:Simulations Parameters}
    \begin{tabular}{l r l r}
    \hline \hline
    Parameters & Values&Parameters & Values \\ \hline \hline
N& 10 - 50 users&$\phi$& 55\% - 95\%\\
K& 8&$r_{0n}$& 1 Mbps\\
B& 20MHz&$N_o$& -170 dBm/Hz\\
$\alpha$& 4&$f_c$& 900 MHz\\
$h_{min},h_{max}$& 21m, 100m&$P^{Max}_t$& 0.2W\\
   \hline
    \end{tabular}}
\end{table}

{\subsubsection{Initialization}
\label{Initialization}
Algorithm \ref{algo:Algorithem 2} works iteratively to find the optimal solution; for this, they need initial values for the inner approximation of \eqref{C3T4}. Furthermore, in an iterative algorithm, the feasibility of the solution is primarily determined by the initialization step. The most common approach to find the initial value for inner approximation is to solve the feasible problem, i.e., \eqref{Transformed3}, without the objective function so that the solution satisfies all of the constraints mentioned. However, it is unattainable because of constraint \eqref{C3T4}, as directly solving \eqref{Transformed3} also requires inner approximation, which is again an iterative process. Therefore, to determine the feasible initial value of $\mathfrak{s}_o$, we solve the left-hand side of \eqref{C3T4} with equal power transmission and by heuristically adjusting the sub-carrier allocation matrix to satisfy the constraints in \eqref{Power allocation},\eqref{Subcarrier 1} and \eqref{Subcarrier 2}. This solution's output is referred to as the initial value of $\mathfrak{s}$. }
\subsubsection{Convergence and Complexity Analysis of Algorithm \ref{algo:Algorithem 2}}

\paragraph{Convergence Analysis}
Denote $\mathbf{\Pi}=\left(\textbf{A}^*,\textbf{P}^*,\mathbf{\eta^*,\mathfrak{s}^*}\right)$ as the optimal solution to problem \eqref{Transformed3} at iteration $t$. Furthermore, the objective functions of the original \eqref{Main Equation3} and approximated \eqref{Transformed3} optimization problem are represented by $\mathbb{F}_1\left(\mathbf{\Pi}\right)$ and $\mathbb{F}_2\left(\mathbf{\Pi}\right)$, respectively. Because the approximated problem's feasible region is a subset of the original optimization problem's feasible region. we have $\mathbb{F}_1\left(\mathbf{\Pi}\right)\ge \mathbb{F}_2\left(\mathbf{\Pi}\right),\forall\mathbf{\Pi}$. Furthermore, the objective function value of 3 improves iteratively, i.e., $\mathbb{F}\left(\mathbf{\Pi}^{t+1}\right) \ge \mathbb{F}\left(\mathbf{\Pi}^{t}\right)$, according to the authors in \cite{beck2010sequential}. The objective function value is bounded by transmission power and QoS constraints and will converge to a local optimal solution. Although it does not guarantee the global optimal solution, where the performance gap to achieve the global optimal solution is left for future work.

\paragraph{Complexity Analysis}
The complexity of Algorithm \ref{algo:Algorithem 2} is determined in this section by considering the number of decision variables and constraints in optimization problem \eqref{Transformed3}. Since optimization problem \eqref{Transformed3} is convex, the interior-point method is applied to find the efficient solution. Specifically,  $3NKL+NL$ decision variables and $8NKL+5NL+KL+L$ constraints in convex problem \eqref{Transformed3} results in the per iteration complexity $\mathcal{O}(\left(3NKL\!+\!NL\right)^3\!\left(8NKL\!+\!5NL\!+\!KL\!+\!L\right))$ \cite{9460776,9526283}. Furthermore, assuming that $t_{m}$ are a total number of iterations, the overall complexity of Algorithm \ref{algo:Algorithem 2} is $\mathcal{O}\left(t_m\!\left(\left(3NKL\!+\!NL\right)^3\!\left(8NKL\!+\!5NL\!+\!KL\!+\!L\!\right)\!\right)\!\right)$.

\section{Results and Discussion}
\label{Results and Discussion}
This section demonstrates the effectiveness of the proposed algorithm via extensive Monte-Carlo simulations using the simulation parameter listed in Table \ref{tab:Simulations Parameters}. In our simulations, the $N$ users are distributed randomly across a $1\text{Km} \times 1\text{Km}$ area. Furthermore, UAVs are deployed to provide communication services using the OFDMA protocol. Depending on the communication environment, Table \ref{tab:modeling parameters} lists additional attenuation coefficients for path loss $\left(\xi_{LoS},\xi_{NLoS} \right)$ and parameters $\left(b_1,b_2\right)$ related to the probability of line-of-sight communication. Furthermore, the performance of the proposed scheme is evaluated for different KPIs, e.g.,  path loss and sum rate. In addition, the numerical results for path loss and sum rate are compared to the respective benchmark schemes.
\par
In the first section of the proposed framework, optimal $3$D placement and user association were achieved by iteratively solving the \eqref{Main EquationP2b3} and \eqref{Main Equation2a}. The goal of path loss minimization is achieved, as described in Section \ref{Proposed Iterative Algorithm PL}, by selecting the $L$ number of active UAVs while taking into account the total number of ground users, network average capacity and users' QoS requirements, e.g., constraint \eqref{C1}. Furthermore, the proposed design is compared with the following references:

\begin{itemize}
    \item \textit{K-Mean Clustering}: This scheme employs the $K$-Mean clustering algorithm to partition the $N$ number of users into $L$ clusters. The location of centroids is regarded as the $2$D location of the UAVs. 

    \item \textit{K-Medioid Clustering}: This scheme is used $K$-Medioid clustering algorithm to cluster the N number of users into the $L$ distant set.

    \item \textit{Mean-Shift Clustering}: The Mean-Shift clustering algorithm determines the $2$D location of UAVs using the Mean-shift algorithm, which works on the principle of kernel density estimation function. 

    \item \textit{Random Association}: In this scheme, $L$ number of  UAVs are optimally deployed using \eqref{Main EquationP2b3}, while users are randomly assigned to the UAVs.
    \item \textit{Random Deployment}: In scheme, $L$ number of UAVs are distributed randomly over an area. whereas the user are allocated to the UAVs using \eqref{Main Equation2a}
\end{itemize}

After determining the optimal 3D location of the UAVs and the users' association matrix, an efficient resource allocation problem \eqref{Transformed3} is solved to achieve the maximum sum rate that satisfies the QoS. Moreover, to demonstrate the \textit{Proposed Scheme's} advantages for sum rate maximization, we compared the results to the following relevant benchmark schemes:
\begin{itemize}
    \item \textit{Ref. \cite{VTCFALL22}}: In this scheme, optimal UAV locations and user association matrices are calculated by solving \eqref{Main EquationP2b3}  and \eqref{Main Equation2a}, respectively, while optimal transmission power and sub-carrier allocation is accomplished by solving \eqref{Transformed3}. Whereas the selection of active UAVs were carried out without considering the QoS constraint of users by using the formula $L=\lceil{\frac{\lambda N}{K}}\rceil$.   
    \item \textit{Equal Power Allocation}: This scheme employs the optimal sub-carrier allocation with an equal transmission power allocation to all users.
    \item \textit{Random Sub-Carrier Allocation}: In this scheme, sub-carriers are allocated at random, while transmission power is allocated optimally.
    \item \textit{Random Power}: This scheme allocates sub-carriers optimally while allocating transmission power randomly to ensure service quality for all users.
    \item \textit{Joint Power and Bandwidth Allocation}\cite{8932190}: In this scheme, A2G communication is carried out over an orthogonal frequency band to minimize inter-cell interference, while UE's sum rate is maximized by optimizing transmission power, bandwidth, and the 3D location of UAV's simultaneously.
\end{itemize}
Furthermore, we perform Monte Carlo simulations, and average results are produced over an independent channel.

\begin{figure}[ht]
	\begin{subfigure}{.5\textwidth}
	\centering
	\includegraphics[width=0.8\linewidth]{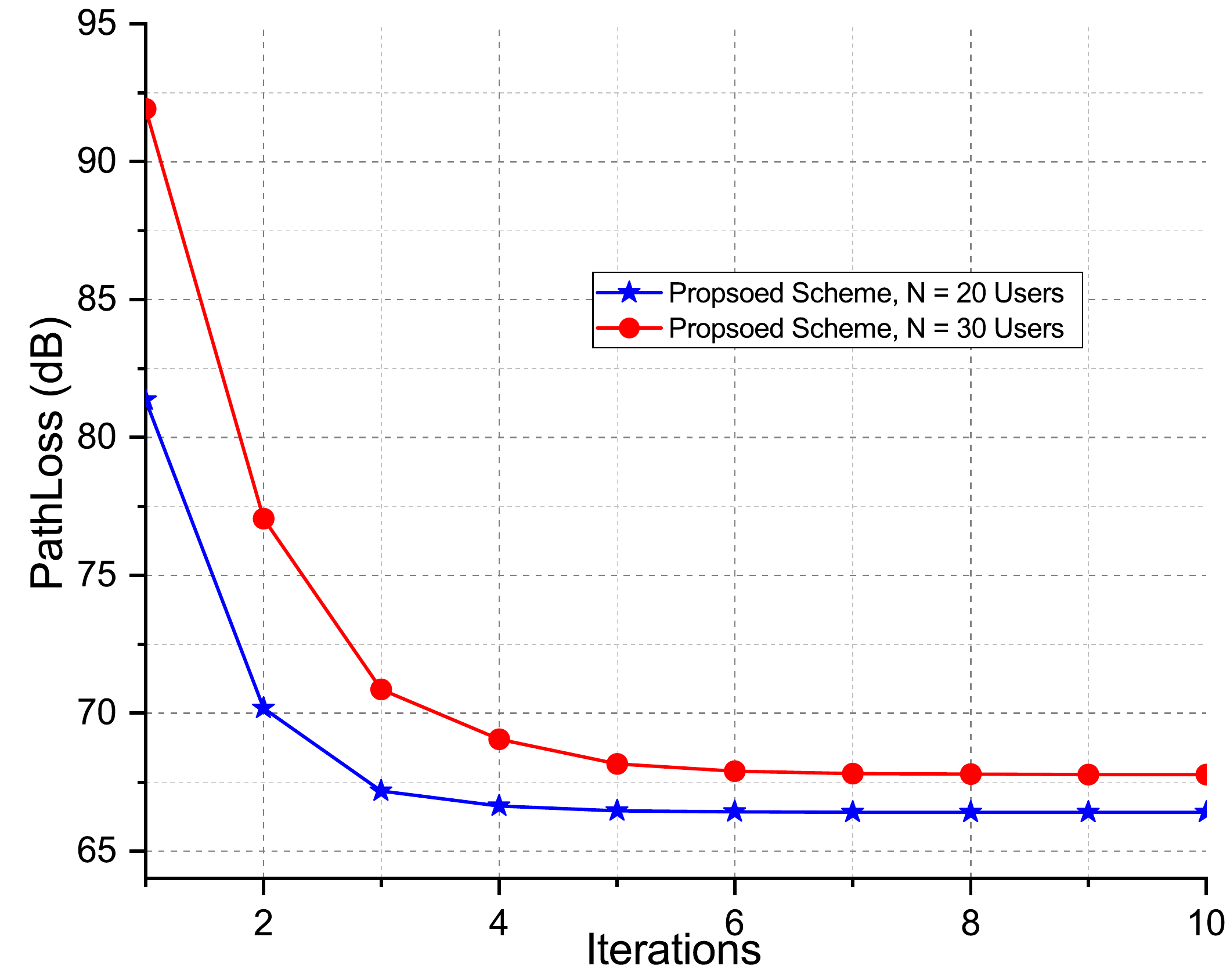}
	\caption{Algorithm 1}
	\label{fig:Algo Pathloss}
	\end{subfigure}\\
	\begin{subfigure}{.5\textwidth}
	\centering
	\includegraphics[width=0.8\linewidth]{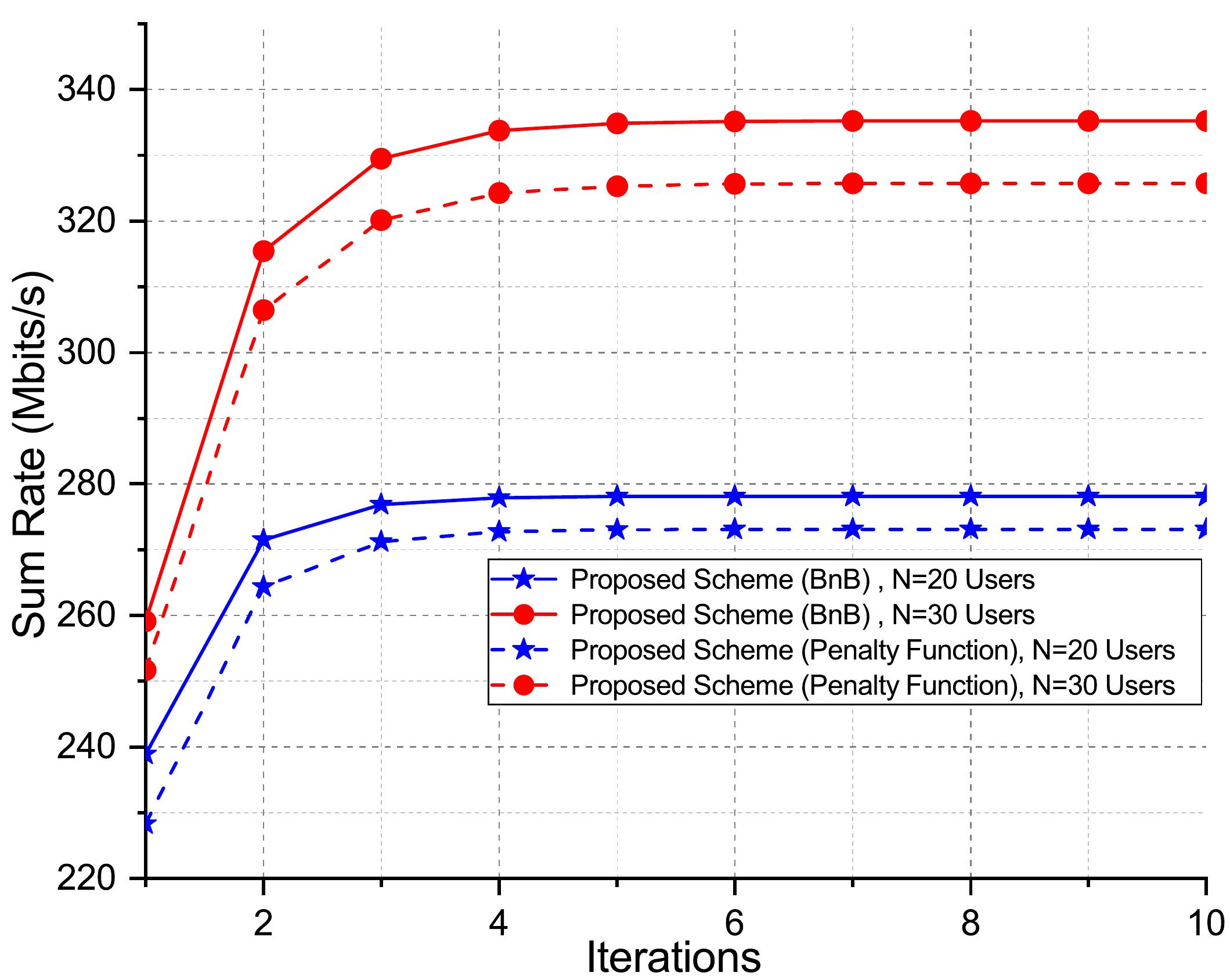}
	\caption{Algorithm 2}
		\label{fig:Algo SumRate}
	\end{subfigure}
	\caption{Convergence Analysis of Algorithm $1$ \& Algorithm $2$}
	\label{fig:Convergance Algo}
\end{figure}

\begin{figure}[ht]
	\centering
	\includegraphics[width=0.8\linewidth]{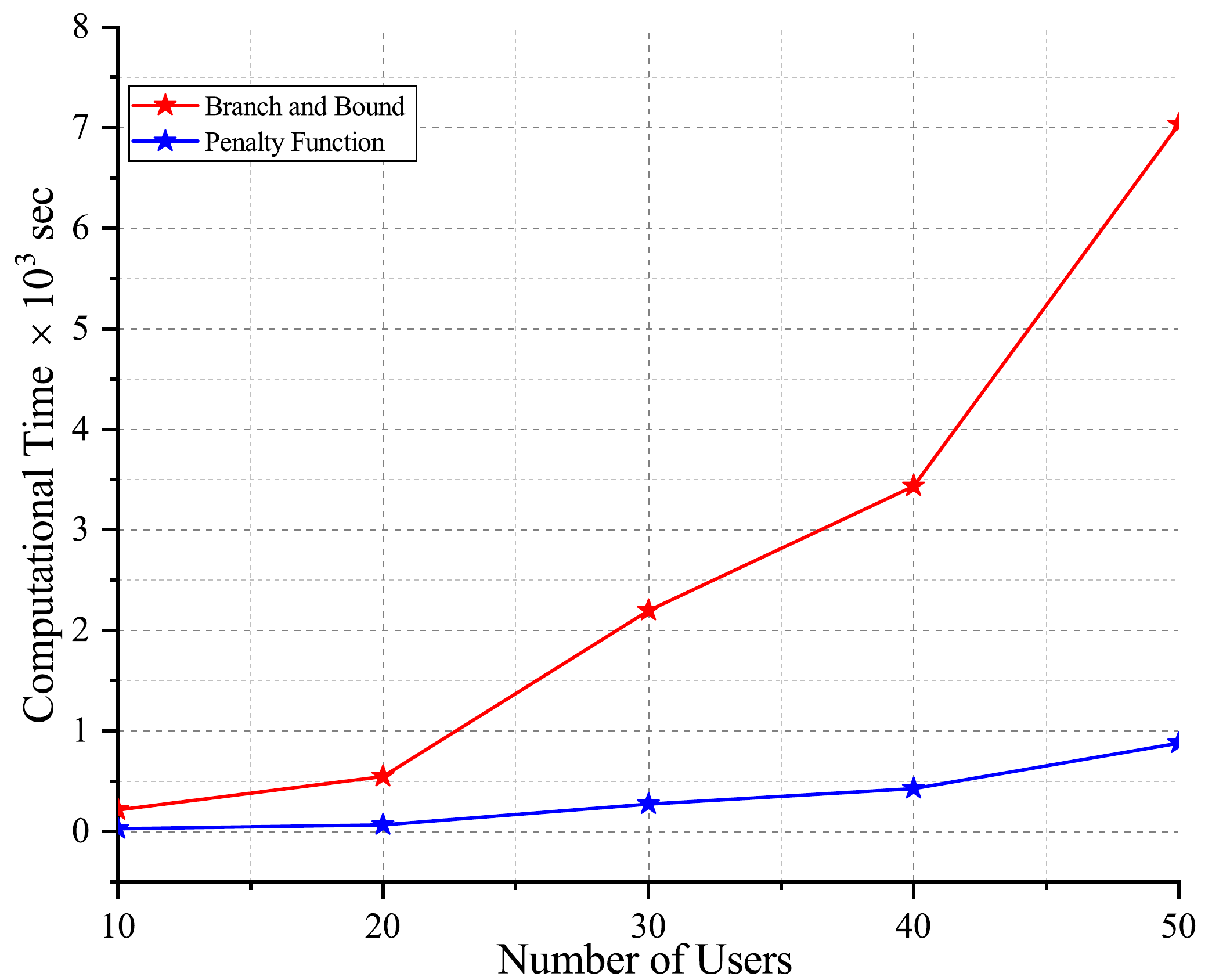}
	\caption{Time Complexity}
	\label{fig:TimeComplexity}
\end{figure}

\subsection{Performance Analysis.}
\label{Performance Analysis}
In Fig.\ref{fig:Convergance Algo}, the convergence of proposed algorithms is demonstrated and compared with different stimulation parameters. Fig.\ref{fig:Algo Pathloss} represents the convergence graph of Algorithm \ref{algo:Algorithem 1} by considering path loss as a performance matrix. Results reveal that as the number of iterations increases, the path loss of the network also decreases. This trend is because UAVs update their locations and user association matrix incrementally. As a result, the distance between UAVs and associated users decreases, and path loss decreases as well. On the other hand, the convergence time for various simulation parameters tends to vary. As the number of users increases, more stimulation parameters are added to the system, requiring more computation to find a feasible solution that minimizes the path loss values. Based on the UAV's optimal $3$D locations and user association matrix solved by Algorithm~\ref{algo:Algorithem 1}, we solve the optimal resource allocation optimization problem using Algorithm \ref{algo:Algorithem 2}. Results in Fig.~\ref{fig:Algo SumRate} represent the convergence behaviour of both \eqref{Transformed3} and \eqref{Transformed3_relaxed} in terms of sum rate as a performance matrix. Results reveal that the algorithm converges to a stable (feasible) point after some iterations.

\par
Fig.~\ref{fig:Algo SumRate} also confirms the effectiveness of the relaxed problem \eqref{Transformed3_relaxed} compared with the mixed binary problem \eqref{Transformed3}, which is revealed via an acceptable tolerance gap between two curves. Although having a slightly degraded performance, the relaxed problem \eqref{Transformed3_relaxed} significantly reduces the computation time compared with the mixed binary problem \eqref{Transformed3}, as shown in Fig. \ref{fig:TimeComplexity}. 
For the small number of users, the time required to compute \eqref{Transformed3} and \eqref{Transformed3_relaxed} is small. In contrast, the computational time required to solve the mixed binary problem grows exponentially with the number of users compared to the relaxed version.
\subsection{Pathloss and Sum Rate Performance}
Because the system sum rate largely depends on the path loss between the UAV and its serving users, we evaluate both path loss and sum rate as the performance matrices of the proposed algorithms. 

\begin{figure}[ht]
	\begin{subfigure}{.5\textwidth}
	\centering
	\includegraphics[width=0.95\linewidth]{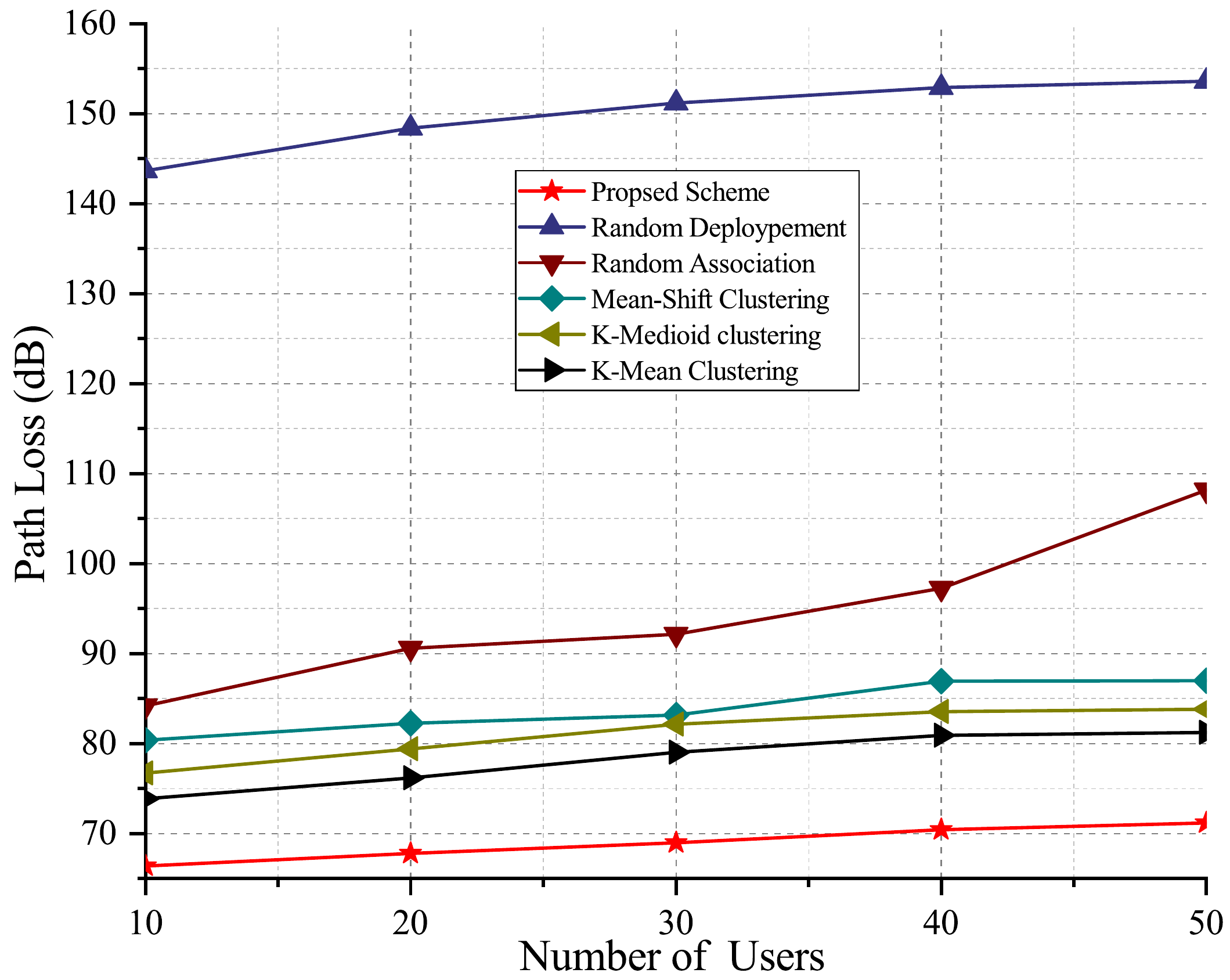}
	\caption{}
	\label{fig:Pathloss_Tech}
	\end{subfigure}\\
	\begin{subfigure}{.5\textwidth}
	\centering
	\includegraphics[width=0.95\linewidth]{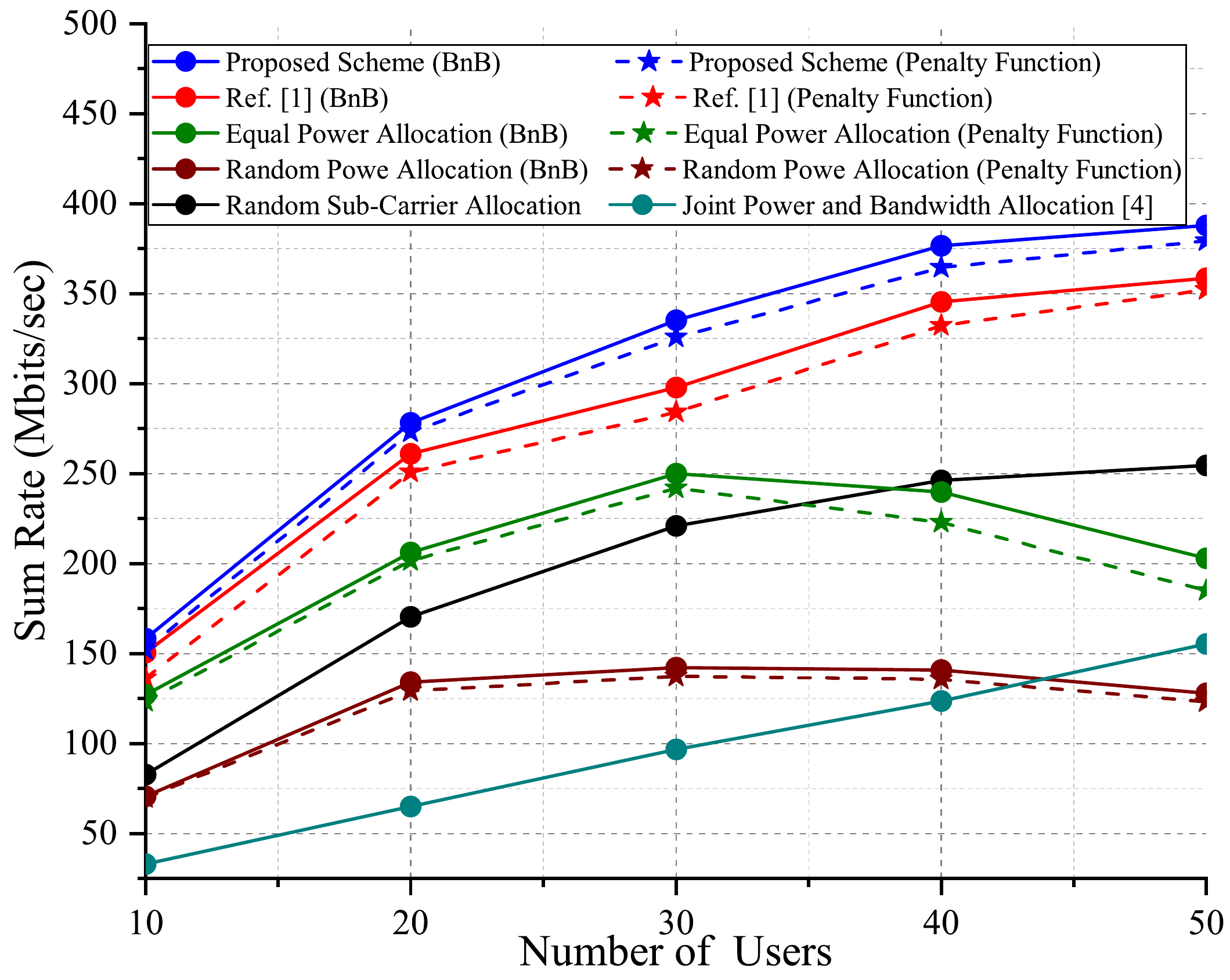}
	\caption{}
		\label{fig:SumRate_Tech}
	\end{subfigure}
	\caption{Comparison of proposed solution}
	\label{fig:Propsed Techniques}
\end{figure}

Fig. \ref{fig:Pathloss_Tech} represents the comparative analysis of the proposed scheme by considering path loss as a performance metric. The outcome reveals that the proposed scheme outperforms all other schemes. In our design, the optimal number of UAVs are selected and placed in the $3$D cartesian coordinate system by considering the user QoS. Likewise, users are connected only to the UAV located at the closest distance compared to others.
\par

Fig. \ref{fig:SumRate_Tech} compares the sum rate of the proposed algorithm with other reference schemes as a function of the users. Superior performance is observed by the proposed algorithm compared with the reference solutions. Furthermore, results reveal a fundamental trade-off between optimal resource allocation and equal power allocation. Results demonstrate that for a small number of users, an equal power allocation scheme performs better than the optimal power allocation; on the other hand, as the number of users increases, the optimal power allocation scheme starts to perform better than others. This trend is because inter-cell interference also increases with the increment in the number of users; as a result, the sum rate starts decreasing in an equal power allocation scheme. 

\begin{figure}[]
	\centering
	\includegraphics[width=0.8\linewidth]{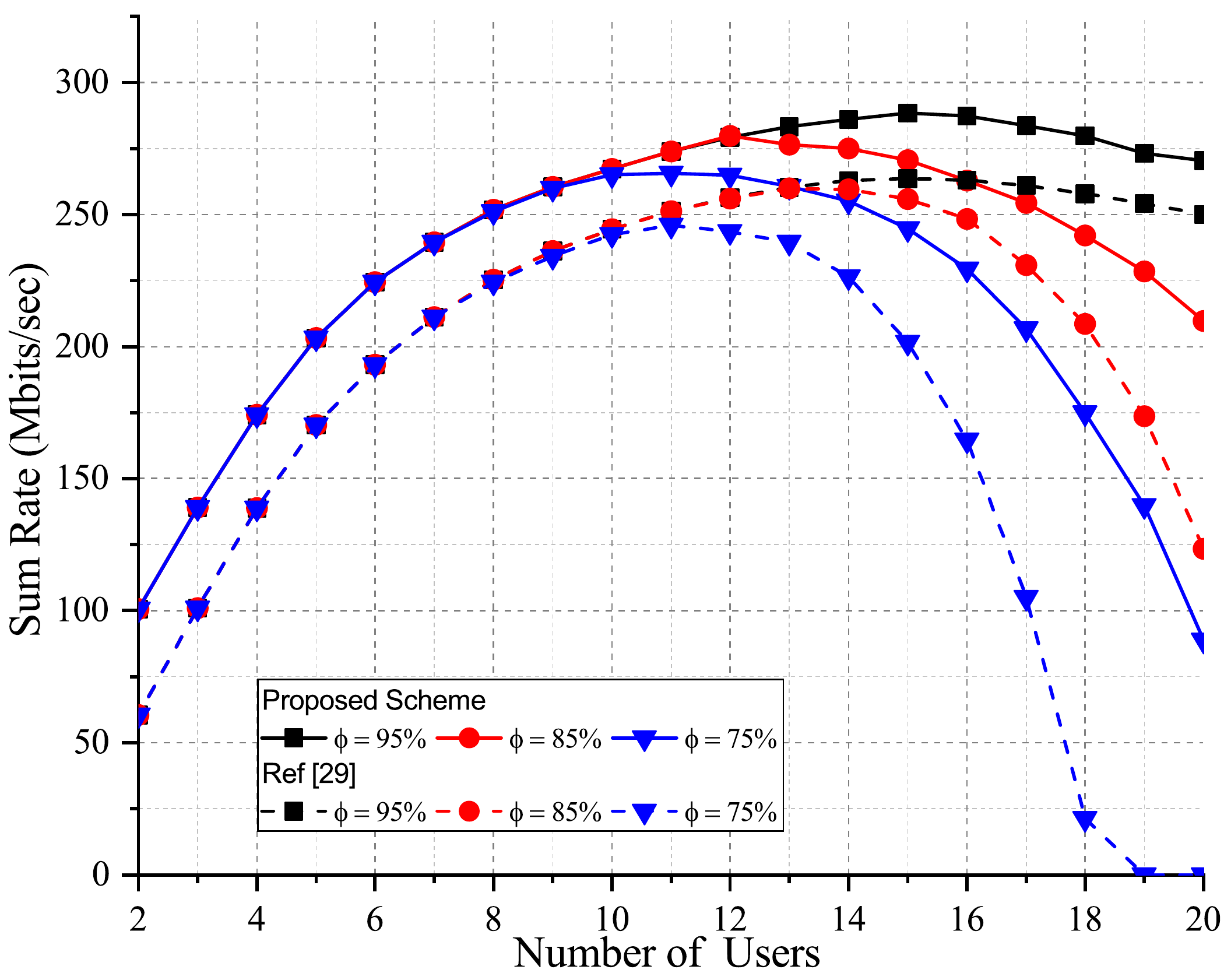}
	\caption{Impact of the probability of line of sight on the system performance.}
	\label{fig:PLOS}
\end{figure}

\begin{figure}[]
	\centering
	\includegraphics[width=0.8\linewidth]{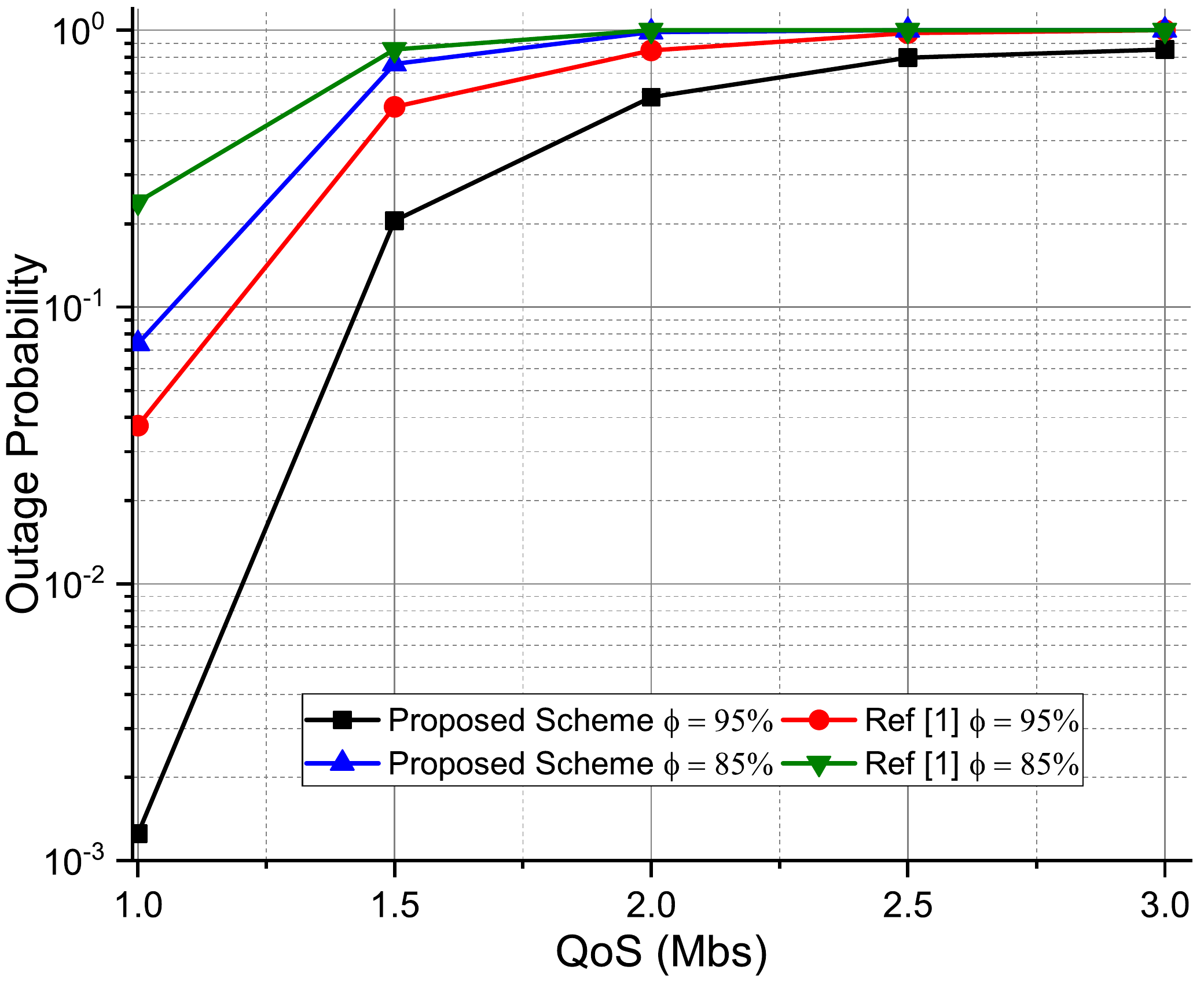}
	\caption{Outage Probability}
	\label{fig:outage}
\end{figure}

\begin{figure}[ht]
	\begin{subfigure}{.5\textwidth}
	\centering
	\includegraphics[width=0.8\linewidth]{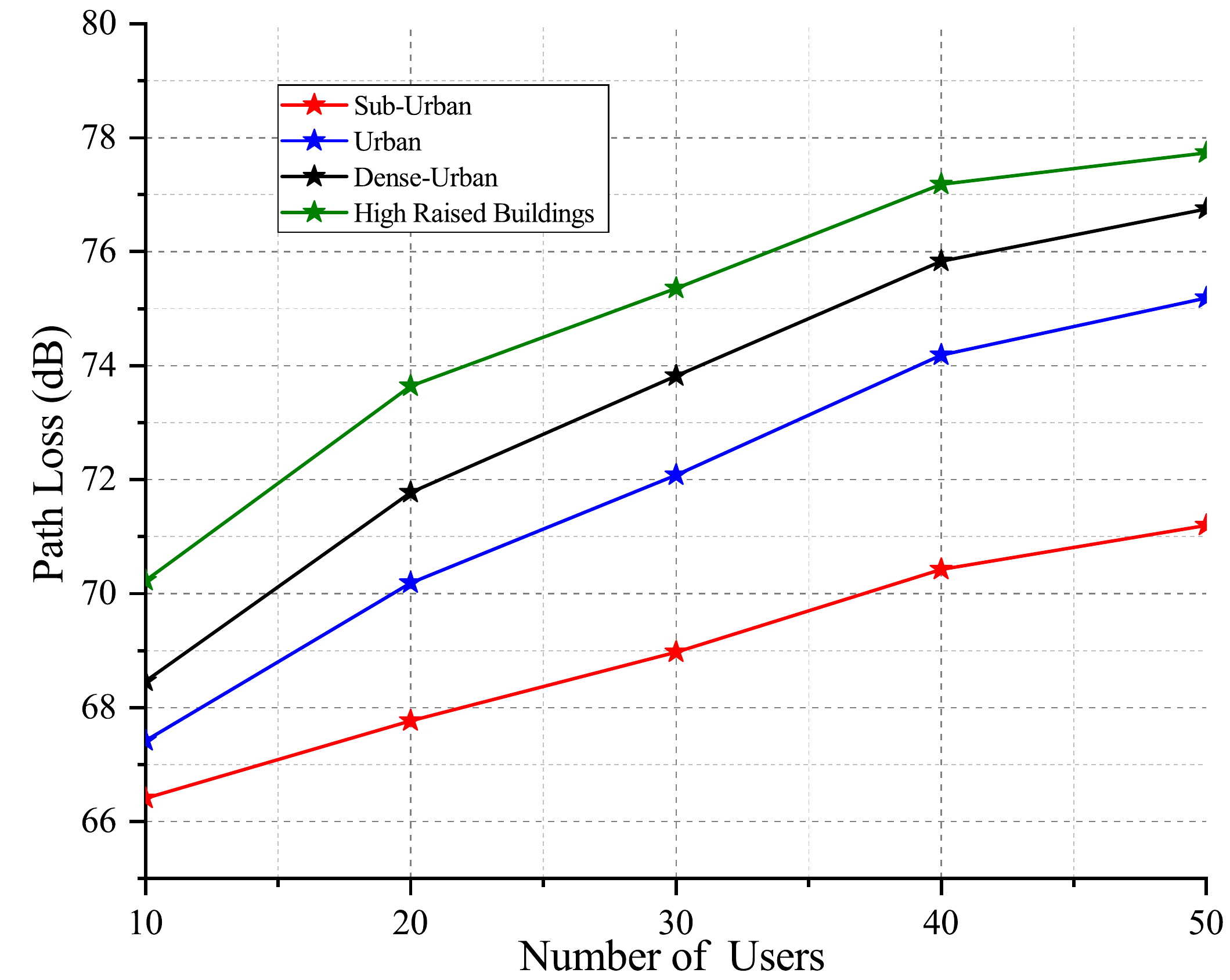}
	\caption{}
	\label{fig:Pathloss_Tech_WE}
	\end{subfigure}\\
	\begin{subfigure}{.5\textwidth}
	\centering
	\includegraphics[width=0.8\linewidth]{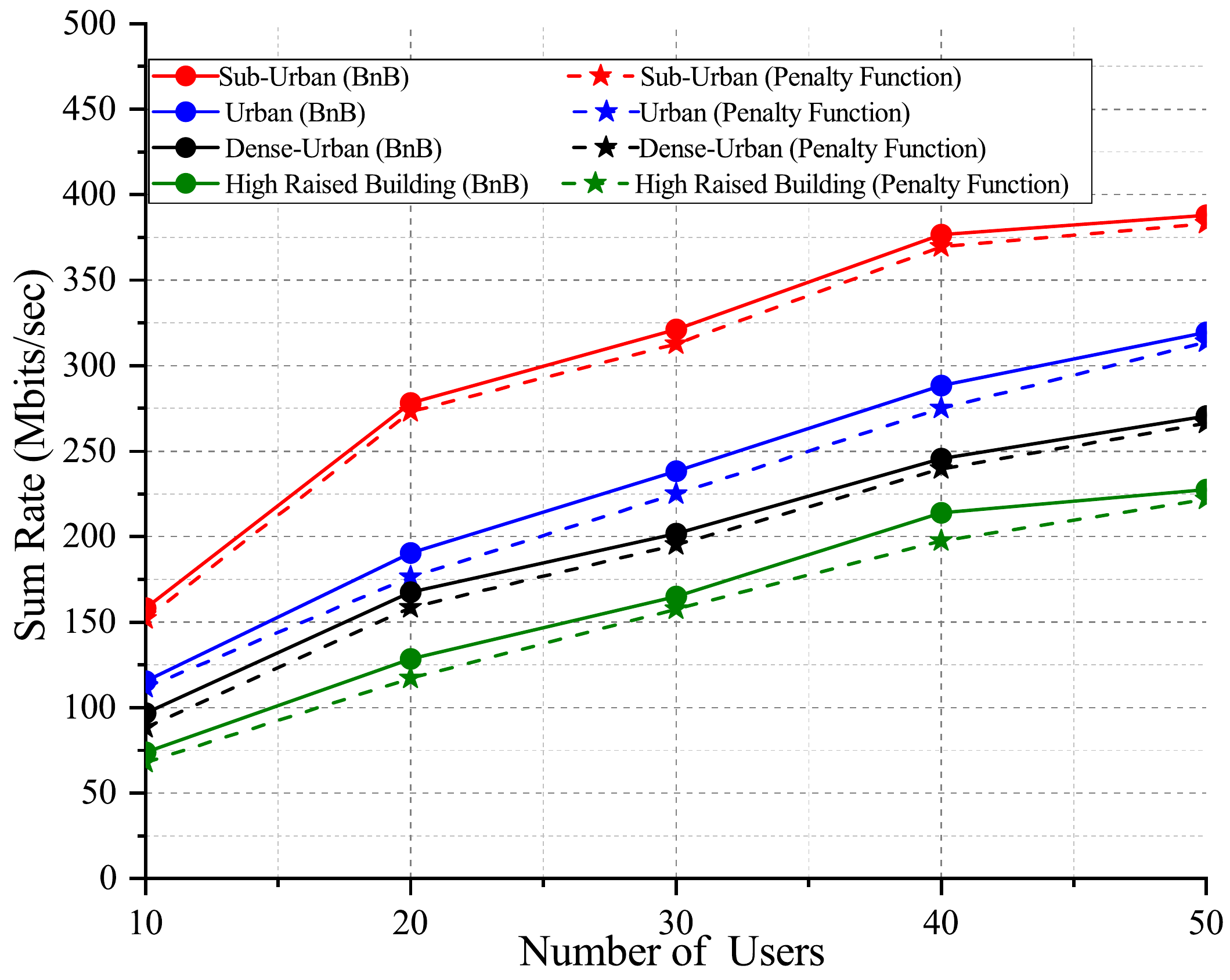}
	\caption{}
		\label{fig:SumRate_Tech_WE}
	\end{subfigure}
	\caption{Impact of working environment on the system performance.}
	\label{fig:Working Environment}
\end{figure}

Similarly, in a UAV communication system, the probability of line-of-sight $(\phi)$ communication is a significant performance-influencing parameter. In practice, not all users have a line-of-sight communication link and have higher path loss than others, hence their QoS compromise. Whereas in our proposed scheme, the selection and deployment of UAVs are carried out by taking into account the QoS requirement of each user and the probability of line of sight, respectively. To analyze the impact, simulations are carried out for different values of  $\phi$, and $N = 20$ users are sorted in ascending order according to path loss as shown in Fig.~\ref{fig:PLOS}. Results reveal that the sum rate increases as the number of users increases.
On the other hand, as there are users with NLoS links, the system sum rate starts decreasing. This trend is because more transmission power is required to satisfy these users' QoS requirements; hence, the sum rate decreases significantly. In the same way, results show that a higher probability of LoS communication results in a higher system sum rate. 
\par

To further reveal the proposed scheme's effectiveness, numerical results are compared with the Ref. \cite{VTCFALL22}, in which the number of UAVs is selected regardless of the user QoS requirements. The results of the Ref. \cite{VTCFALL22} for different $\phi$ demonstrate that at $\phi= 95\%$, Ref. \cite{VTCFALL22} can accommodate the number of non-line of sight users. In contrast, for $\phi= {85,75}\%$ number of non-line of sight users are more significant as compared to $\phi= 95\%$; hence UAVs do not have enough required transmission power to accommodate them; as a result, sum rate decrease significantly for $\phi=85\%$ and goes to zero for $\phi=75\%$ respectively.
The proposed approach's significance in comparison to the theoretical minimum approach can be seen more explicitly by considering outage probability as a performance metric across varying spectral efficiency, as shown in Fig. ~\ref{fig:outage}. The findings demonstrate the effectiveness of the proposed approach in comparison to others.
\par

Similarly, simulations were carried out to further analyze the performance for path loss and sum rate problems in suburban, urban, dense-urban, and high-raised building environments. Results in Fig.~\ref{fig:Pathloss_Tech_WE} demonstrate that the sub-urban environment constitutes minimal path loss. On the other hand, as we move toward other working environments, values of the additional coefficient for path loss increase as given in Table \ref{tab:modeling parameters}. Following that, the system's performance is further evaluated for the sum rate performance metric under different environments, as shown in Fig.~\ref{fig:SumRate_Tech_WE}. Results reveal that the system sum rate is higher for a sub-urban working environment than others because of its minimal path loss. Similarly, Fig. ~\ref{fig:Working Environment} illustrates the working environment's impact on the system's performance.
\par
\begin{figure}[ht]
	\begin{subfigure}{.5\textwidth}
	\centering
	\includegraphics[width=0.85\linewidth]{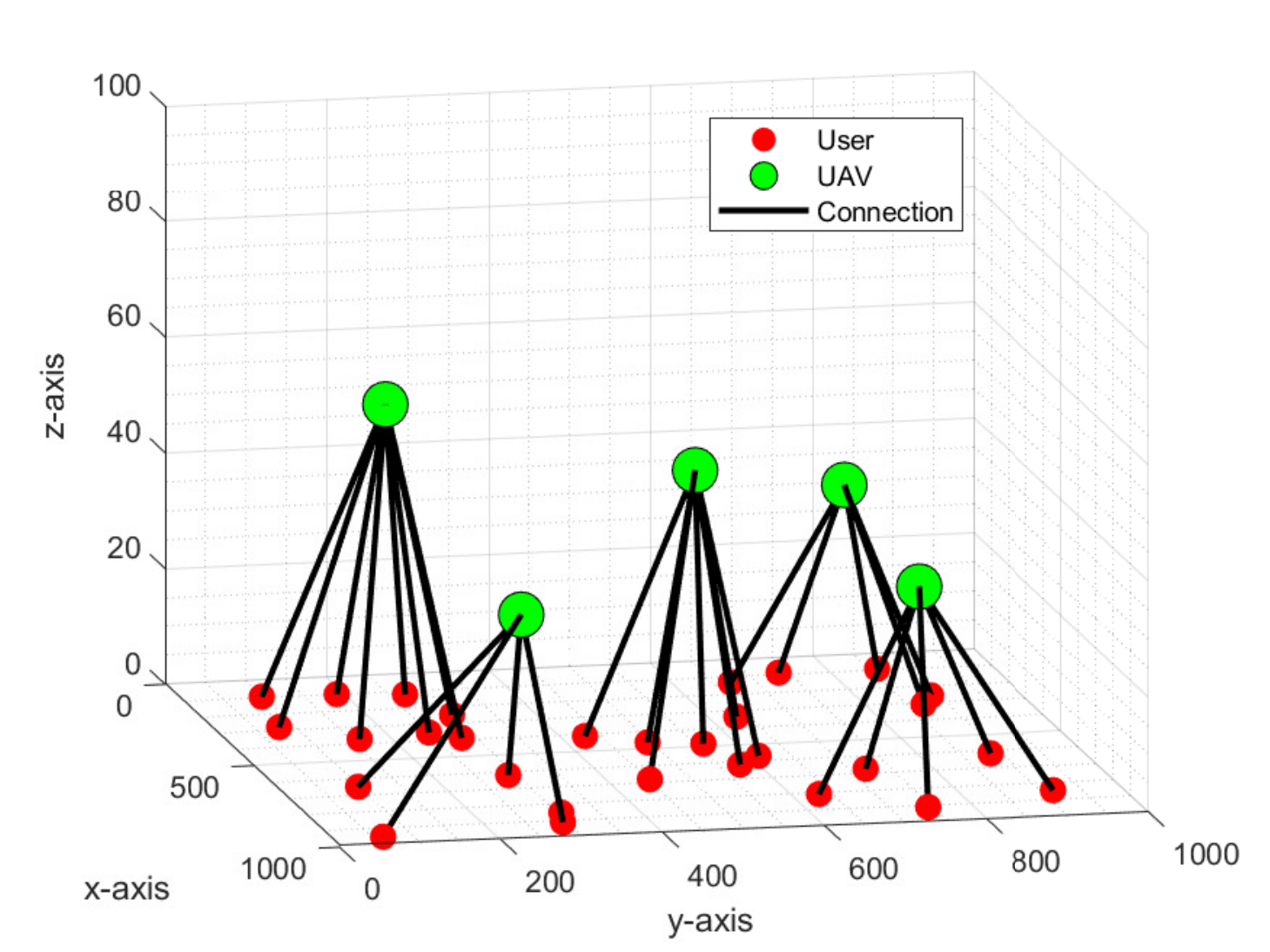}
	\caption{UAV-UE connectivity}
	\label{fig:U=30 Association}
	\end{subfigure}\\
	\begin{subfigure}{0.5\textwidth}
	\centering
	\includegraphics[width=0.8\linewidth]{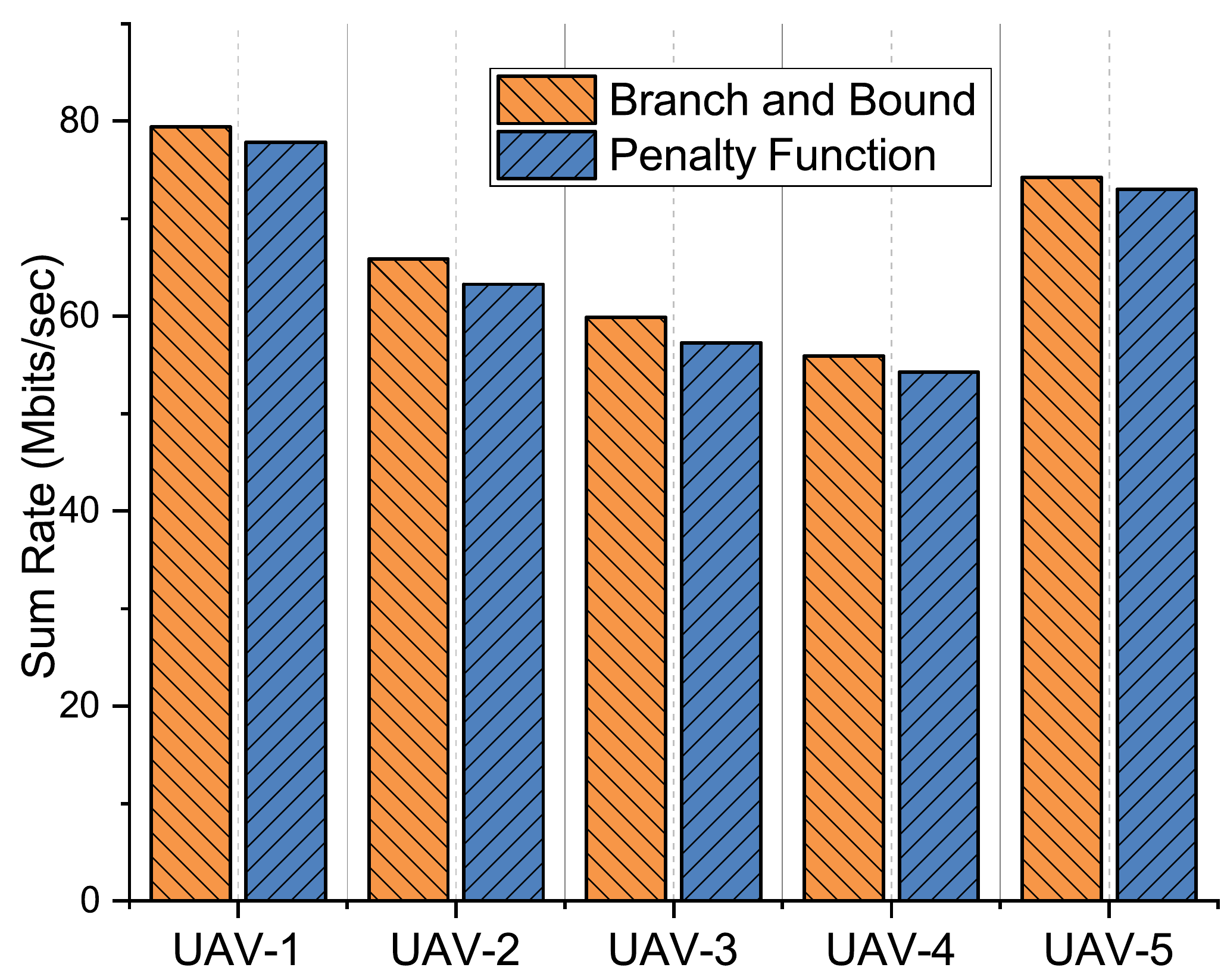}
	\caption{Rate per UAV}
		\label{fig:RatePerUAV}
	\end{subfigure}
	\caption{UAV-UE connectivity and individual UAV rate for N=30 .}
	\label{fig:User Association}
\end{figure}

{Next, the proposed scheme performance is further validated by taking into account the UAV-UE connectivity and individual sum rate of different UAVs for $N$$=$$30$ as shown in Fig. \ref{fig:User Association}. In particular, Fig. \ref{fig:U=30 Association} shows UAV-UE connectivity, where the proposed scheme requires 5 UAVs to serve 30 UEs. According to the proposed scheme, UAVs are deployed optimally at a point such that the distance between UAVs and associated users is minimized}. Moreover, the impact of the proposed association scheme can be seen in Fig.\ref{fig:RatePerUAV} by plotting the sum rate of individual UAVs. It can be observed that our approach produces a load-balanced system configuration, and there is no significant deviation among the load of different UAVs.

\section{Conclusions and Discussions}
\label{Conclusion}
In this work, we have considered an ODFMA-enabled multiple UAV communication system to provide on-demand communication services in emergency scenarios. In this context, We formulate a noval problem to maximize the sum rate via a joint optimization of the number of active UAVs, their 3D placement, and the radio resource allocation under a limited transmit power budget and users' QoS requirements. Since the original optimization problem is not convex, we first decoupled it into sub-problems, which we then solved iteratively to find the best feasible solution. Furthermore, the effectiveness of the proposed scheme is illustrated via numerical results. Results demonstrate that the proposed scheme outperforms the benchmark schemes in terms of both path loss and user sum rate. We showed that optimal deployment and user association minimize path loss, directly impacting the system sum rate. The results also demonstrated that by increasing the probability of the light-of-sight channel, the path loss decreases significantly, thereby improving the system sum rate.

\par
The outcomes of this work suggest several promising research topics. The first topic is to consider the user dynamic, in which the number of users is time-varying. In such cases, the number of active UAVs should be planned jointly with the prediction of active users. The second topic is studying the UAV movement in the deployment design. This may reduce the number of active UAVs; however, the problem will be more challenging.

\begin{appendices}
\section{Proof of LEMMA 1}\label{Lemma1}
We can prove the expressions $\textbf{H}_o$,  $\textbf{H}_n$, $\textbf{F}_o$, $\textbf{F}_n$, $\kappa_o$ and $\kappa_n$ by considering the objective function and constraints of \eqref{Main Equation2b}, which are represented by $f_1$ and $f_2$ respectively:
\begin{equation}
\small
\label{OT}
   f_1(\mathbf{W}) ={\sum}_{n = 1}^{{Q^l}} {\left[ {{{\left({{x_n} - {x_l}} \right)}^2} + {{\left( {{y_n} - {y_l}} \right)}^2} + h_l^2} \right],\forall l} . 
\end{equation}
\vspace{-3mm}
\begin{equation}
\small
\label{CT}
   f_2(\mathbf{W})={\left( {{x_n} - {x_l}} \right)^2} + {\left( {{y_n} - {y_l}} \right)^2} + h_l^2\xi  \le \rm{0} .
\end{equation}
In \eqref{Main EquationP2b3}, $\mathbf{H_o}$,  and $\mathbf{H_n}$ represent the positive semi-definite matrices that can be calculated by using the second partial derivatives of $f_1$ and $f_2$, respectively.
As a result, the Hessian of $f_1(\mathbf{W})$ and $f_2(\mathbf{W})$ can be defined as follows:
\vspace{-2mm}
	\begin{equation}
	\small
	\label{H_o}
	\mathbf{H_o}=	{\nabla ^2}f_1(\mathbf{W}) = \left[ {\begin{array}{*{20}{c}}
				2Q^l&0&0\\
				0&2Q^l&0\\
				0&0&2Q^l
		\end{array}} \right]=2Q^l\mathbf{I_3}.	
	\end{equation}
	\begin{equation}
	\small
	\label{H_n}
	\mathbf{H_n}=	{\nabla ^2}f(\mathbf{W}) = \left[ {\begin{array}{*{20}{c}}
				2&0&0\\
				0&2&0\\
				0&0&\xi
		\end{array}} \right].		
	\end{equation}
Equation \eqref{H_o} and \eqref{H_n} represents the Hessian matrix of $f_1$ and $f_2$, whose all the elements are zeros except the diagonal elements such that $\left| {{\nabla ^2}(f_1(\mathbf{W}))} \right| > 0$ and $\left| {{\nabla ^2}(f_2(\mathbf{W}))} \right| > 0$. As a result, the objective function and constraints $\textbf{C1}$ specified in \eqref{Main Equation2b} are convex in nature.
\par
Following that, the transformation from \eqref{Main Equation2b} to \eqref{Main EquationP2b3} can be proven by simply expanding and rearranging the quadratic terms in \eqref{OT} and \eqref{CT} in the following manner:
\vspace{-2mm}
\begin{equation}
\small
\begin{aligned}
\label{OT1_3}
     f_1(\mathbf{W})\!\! =\!\!\underbrace{{\sum}_{n = 1}^{Q^l}\!\!\left({x^l}^2\!\!\!\!+\!{y^l}^2\!\!\!\!+\!{h^l}^2\right)}_{\text{OT1}}\!+\!\!\underbrace{{\sum}_{n = 1}^{Q^l}\!\!\left(\!-\!2x_n{x^l}\!\!\!-\!2y_n{y^l}\right)}_{\text{OT2}}\\
     \!+\!
     \underbrace{{\sum}_{n = 1}^{Q^l}\!\!\left(\!x_n^2\!+\!y_n^2\right)}_{\text{OT3}}.
\end{aligned}
\end{equation}
\vspace{-2mm}
Similarly,

\begin{equation}
\small
\label{CT1_3}
\begin{aligned}
     f_2(\mathbf{W})\!\! =\!\!\underbrace{\left({x^l}^2\!\!\!\!+\!{y^l}^2\!\!\!\!+\!\xi{h^l}^2\right)}_{\text{CT1}}\!+\!\!\underbrace{\left(\!-\!2x_n{x^l}\!\!\!-\!2y_n{y^l}\right)}_{\text{CT2}}
     \!+\!
     \underbrace{\left(x_n^2\!+\!y_n^2\right)}_{\text{CT3}}.
\end{aligned}
\end{equation}
To prove the transformation of \eqref{P2b Objective} to \eqref{EquationP2b2 Objective}, we start solving \eqref{OT1_3} and find the value of $\text{OT1}$, $\text{OT2}$ and $\text{OT3}$ respectively.  As perceived from \eqref{OT1_3} we have
\vspace{-3mm}
\begin{equation}
\small
\label{OT1}
\text{OT1}={\sum}_{n = 1}^{Q^l}\!\!\left({x^l}^2\!\!\!\!+\!{y^l}^2\!\!\!\!+\!{h^l}^2\right),     
\end{equation}
whereas ${x^l}^2\!\!\!\!+\!{y^l}^2\!\!\!\!+\!{h^l}^2$ in matrix notation can be represented as follows:
\begin{equation}
\small
{x^l}^2\!\!\!\!+\!{y^l}^2\!\!\!\!+\!{h^l}^2=
\begin{bmatrix}
\label{OT1_a}
x^l\\y^l\\h^l
\end{bmatrix}^T
\begin{bmatrix}
1&0&0\\
0&1&0\\
0&0&1
\end{bmatrix}
\begin{bmatrix}
x^l\\y^l\\h^l
\end{bmatrix}=\mathbf{W}^T\mathbf{I_3}\mathbf{W},
\end{equation}
Putting the value of \eqref{OT1_a} in \eqref{OT1}, we have \vspace{-3mm}
\begin{equation}
\small
\vspace{-2mm}
\label{OT1_b}
    \text{OT1}={\sum}_{n = 1}^{Q^l}\mathbf{W}^T\mathbf{I_3}\mathbf{W}=\mathbf{W}^TQ^l\mathbf{I_3}\mathbf{W},
\end{equation}
Whereas, form \eqref{H_o}, $Q^l\mathbf{I_3}=\frac{1}{2}\mathbf{H_o}$ and
by putting the value in \eqref{OT1_b}, we have
\vspace{-3mm}
\begin{equation}
\small
\label{OT1_f}
    \text{OT1}=\frac{1}{2}\mathbf{W}^T\mathbf{H_o}\mathbf{W}.
\end{equation}
Similarly from \eqref{OT1_3}, $\text{OT2}$ in matrix notation can be represented as follows:
\vspace{-2mm}
\begin{equation}
\small
\label{OT2}
    \text{OT2}=\!\sum\limits_{n = 1}^{Q^l}\!\!\left(\!-\!2x_n{x^l}\!\!\!-\!2y_n{y^l}\right)\!=\!\!\underbrace{\begin{bmatrix}
    -2\!\!\sum\limits_{n = 1}^{Q^l}\!x_n\!\\-2\!\!\sum\limits_{n = 1}^{Q^l}\!y_n\!\\0
    \end{bmatrix}^T}_{\mathbf{F}_o}
    \!\!\!\!\!\!\begin{bmatrix}
    x^l\\y^l\\h^l
    \end{bmatrix}\!=\!\mathbf{F}_o^T\mathbf{W}.
\end{equation}
Similarly from \eqref{OT1_3}, $\text{OT3}$ represented as 
\vspace{-2mm}
\begin{equation}
\small
    \label{OT3}
    \text{OT3}={\sum}_{n = 1}^{Q^l}\left(x_n^2\!+\!y_n^2\right)=\kappa_o.
\end{equation}
Following the same procedure, we find the value of $\text{CT1}=\frac{1}{2}\mathbf{W}^T\mathbf{H_n}\mathbf{W}$, $\text{CT2} = \mathbf{F}_n\mathbf{W}$ and $\text{CT3}= \kappa_n=x_n^2+y_n^2$.

Putting the value of $\text{OT1}$, $\text{OT2}$ and $\text{OT3}$ in \eqref{OT1_3} and $\text{CT1}$, $\text{CT2}$ and $\text{CT3}$ in \eqref{CT1_3} we have 
  $f_1(\mathbf{W})=\frac{1}{2}{{\mathbf{{W}}}^T}{\textbf{H} _o}\mathbf{W} + \mathbf{F}_o^T{\textbf{W} }+{\kappa_o}$
  and 
 $ f_2(\mathbf{W})=\frac{1}{2}{{\mathbf{{W}}}^T}{\textbf{H} _n}\mathbf{W} + \mathbf{F}_n^T{\mathbf{W} }+{\kappa _n}$, 

which completes the proof.
\end{appendices}
\bibliographystyle{IEEEtran}
\bibliography{ReferenceBibFile}

\begin{thebibliography}{10}
\providecommand{\url}[1]{#1}
\csname url@samestyle\endcsname
\providecommand{\newblock}{\relax}
\providecommand{\bibinfo}[2]{#2}
\providecommand{\BIBentrySTDinterwordspacing}{\spaceskip=0pt\relax}
\providecommand{\BIBentryALTinterwordstretchfactor}{4}
\providecommand{\BIBentryALTinterwordspacing}{\spaceskip=\fontdimen2\font plus
\BIBentryALTinterwordstretchfactor\fontdimen3\font minus
  \fontdimen4\font\relax}
\providecommand{\BIBforeignlanguage}[2]{{%
\expandafter\ifx\csname l@#1\endcsname\relax
\typeout{** WARNING: IEEEtran.bst: No hyphenation pattern has been}%
\typeout{** loaded for the language `#1'. Using the pattern for}%
\typeout{** the default language instead.}%
\else
\language=\csname l@#1\endcsname
\fi
#2}}
\providecommand{\BIBdecl}{\relax}
\BIBdecl

\bibitem{VTCFALL22}
A.~Mahmood, T.~X. Vu, S.~K. Sharma, S.~Chatzinotas, and B.~Ottersten,
  ``Multi-objective optimization for 3d placement and resource allocation in
  ofdma-based multi-uav networks,'' \emph{arXiv preprint arXiv:2304.12798},
  2023.

\bibitem{8744514}
A.~A. {Khuwaja} \emph{et~al.}, ``Optimum deployment of multiple \text{UAV}s for
  coverage area maximization in the presence of co-channel interference,''
  \emph{IEEE Access}, vol.~7, pp. 85\,203--85\,212, 2019.

\bibitem{7510820}
R.~I. Bor-Yaliniz \emph{et~al.}, ``Efficient \text{3-D} placement of an aerial
  base station in next generation cellular networks,'' in \emph{2016 IEEE
  International Conference on Communications (ICC)}, 2016, pp. 1--5.

\bibitem{8932190}
P.~Li and J.~Xu, ``Placement optimization for \text{UAV}-enabled wireless
  networks with multi-hop backhauls,'' \emph{Journal of Communications and
  Information Networks}, vol.~3, no.~4, pp. 64--73, 2018.

\bibitem{8701700}
A.~Masaracchia \emph{et~al.}, ``An energy-efficient clustering and routing
  framework for disaster relief network,'' \emph{IEEE Access}, vol.~7, pp.
  56\,520--56\,532, 2019.

\bibitem{9210567}
O.~Kodheli \emph{et~al.}, ``Satellite communications in the new space era: A
  survey and future challenges,'' \emph{IEEE Communications Surveys \&
  Tutorials}, vol.~23, no.~1, pp. 70--109, 2021.

\bibitem{7881122}
E.~Kalantari \emph{et~al.}, ``On the number and \text{3D} placement of drone
  base stations in wireless cellular networks,'' in \emph{2016 IEEE 84th
  Vehicular Technology Conference (VTC-Fall)}, 2016, pp. 1--6.

\bibitem{8918497}
Y.~Zeng \emph{et~al.}, ``Accessing from the sky: A tutorial on \text{UAV}
  communications for \text{5G} and beyond,'' \emph{Proceedings of the IEEE},
  vol. 107, no.~12, pp. 2327--2375, 2019.

\bibitem{8660516}
M.~{Mozaffari} \emph{et~al.}, ``A tutorial on \text{UAV}s for wireless
  networks: Applications, challenges, and open problems,'' \emph{IEEE
  Communications Surveys Tutorials}, vol.~21, no.~3, pp. 2334--2360, 2019.

\bibitem{7417609}
M.~Mozaffari \emph{et~al.}, ``Drone small cells in the clouds: Design,
  deployment and performance analysis,'' in \emph{2015 IEEE Global
  Communications Conference (GLOBECOM)}, 2015, pp. 1--6.

\bibitem{9627548}
Y.~Liang \emph{et~al.}, ``Joint trajectory and resource optimization for
  \text{UAV}-aided two-way relay networks,'' \emph{IEEE Transactions on
  Vehicular Technology}, vol.~71, no.~1, pp. 639--652, 2022.

\bibitem{8489918}
L.~Xie, J.~Xu, and R.~Zhang, ``Throughput maximization for \text{UAV}-enabled
  wireless powered communication networks,'' \emph{IEEE Internet of Things
  Journal}, vol.~6, no.~2, pp. 1690--1703, 2019.

\bibitem{8618602}
G.~Zhang, Q.~Wu, M.~Cui, and R.~Zhang, ``Securing \text{UAV} communications via
  joint trajectory and power control,'' \emph{IEEE Transactions on Wireless
  Communications}, vol.~18, no.~2, pp. 1376--1389, 2019.

\bibitem{9422153}
S.~Zeng \emph{et~al.}, ``Trajectory optimization and resource allocation for
  ofdma \text{UAV} relay networks,'' \emph{IEEE Transactions on Wireless
  Communications}, vol.~20, no.~10, pp. 6634--6647, 2021.

\bibitem{7918510}
M.~{Alzenad} \emph{et~al.}, ``\text{3-D} placement of an unmanned aerial
  vehicle base station (\text{UAV}-bs) for energy-efficient maximal coverage,''
  \emph{IEEE Wireless Comm. Letters}, vol.~6, pp. 434--437, 2017.

\bibitem{8698468}
C.~You and R.~Zhang, ``\text{3D} trajectory optimization in rician fading for
  \text{UAV}-enabled data harvesting,'' \emph{IEEE Transactions on Wireless
  Communications}, vol.~18, no.~6, pp. 3192--3207, 2019.

\bibitem{8821282}
F.~{Huang} \emph{et~al.}, ``Multiple-\text{UAV}-assisted swipt in internet of
  things: User association and power allocation,'' \emph{IEEE Access}, vol.~7,
  pp. 124\,244--124\,255, 2019.

\bibitem{8982086}
L.~Xie \emph{et~al.}, ``Common throughput maximization for \text{UAV}-enabled
  interference channel with wireless powered communications,'' \emph{IEEE
  Transactions on Communications}, vol.~68, no.~5, pp. 3197--3212, 2020.

\bibitem{9442809}
Z.~Rahimi \emph{et~al.}, ``An efficient \text{3-D} positioning approach to
  minimize required \text{UAV}s for iot network coverage,'' \emph{IEEE Internet
  of Things Journal}, vol.~9, no.~1, pp. 558--571, 2022.

\bibitem{8944019}
R.~{Chen}, X.~{Li}, Y.~{Sun}, S.~{Li}, and Z.~{Sun}, ``Multi-\text{UAV}
  coverage scheme for average capacity maximization,'' \emph{IEEE
  Communications Letters}, vol.~24, no.~3, pp. 653--657, 2020.

\bibitem{6824752}
J.~G. {Andrews} \emph{et~al.}, ``What will \text{5G} be?'' \emph{\small{IEEE
  Journal on Selected Areas in Comm}}, vol.~32, no.~6, pp. 1065--1082, 2014.

\bibitem{8727504}
X.~{Liu} \emph{et~al.}, ``Trajectory design and power control for
  multi-\text{UAV} assisted wireless networks: A machine learning approach,''
  \emph{IEEE Tran. on Vehicular Technology}, vol.~68, pp. 7957--7969, 2019.

\bibitem{8833519}
H.~El~Hammouti, M.~Benjillali, B.~Shihada, and M.-S. Alouini,
  ``Learn-as-you-fly: A distributed algorithm for joint \text{3D} placement and
  user association in multi-{\text{uav}}s networks,'' \emph{IEEE Transactions
  on Wireless Communications}, vol.~18, no.~12, pp. 5831--5844, 2019.

\bibitem{7037248}
A.~Al-Hourani \emph{et~al.}, ``Modeling air-to-ground path loss for low
  altitude platforms in urban environments,'' in \emph{2014 IEEE Global
  Communications Conference}, 2014, pp. 2898--2904.

\bibitem{lu2011kkt}
C.~Lu, S.-C. Fang, Q.~Jin, Z.~Wang, and W.~Xing, ``\text{KKT} solution and
  conic relaxation for solving quadratically constrained quadratic programming
  problems,'' \emph{SIAM Journal on Optimization}, vol.~21, no.~4, pp.
  1475--1490, 2011.

\bibitem{lobo1998applications}
M.~S. Lobo, L.~Vandenberghe, S.~Boyd, and H.~Lebret, ``Applications of
  second-order cone programming,'' \emph{Linear algebra and its applications},
  vol. 284, no. 1-3, pp. 193--228, 1998.

\bibitem{beck2010sequential}
A.~Beck \emph{et~al.}, ``A sequential parametric convex approximation method
  with applications to nonconvex truss topology design problems,''
  \emph{Journal of Global Optimization}, vol.~47, no.~1, pp. 29--51, 2010.

\bibitem{9460776}
T.~S. Abdu \emph{et~al.}, ``Flexible resource optimization for \text{GEO}
  multibeam satellite communication system,'' \emph{IEEE Transactions on
  Wireless Communications}, vol.~20, no.~12, pp. 7888--7902, 2021.

\bibitem{9526283}
T.~X. Vu \emph{et~al.}, ``Dynamic bandwidth allocation and precoding design for
  highly-loaded multiuser $\text{MISO}$ in beyond \text{5G} networks,''
  \emph{IEEE Transactions on Wireless Communications}, vol.~21, no.~3, pp.
  1794--1805, 2022.

\end{thebibliography}
\end{document}